\documentclass[epj]{svjour}
\usepackage{graphics}
\usepackage{graphicx}%
\usepackage{xcolor}%
\begin{document}
\title{AAK24: Global QCD analysis on polarized parton distribution in the presence of $A_2$ asymmetry measurements}
\author{Fatemeh Arbabifar\inst{1}\thanks{\emph {F.Arbabifar@cfu.ac.ir} } \and Shahin Atashbar Tehrani\inst{2,3}\thanks{\emph {Atashbar@ipm.ir} } \and Hamzeh Khanpour\inst{4,2,5}\thanks{\emph {Hamzeh.Khanpour@cern.ch} } 
%
}                     
%
%
\institute{Department of Physics Education, Farhangian University, P.O.Box 14665-889, Tehran, Iran. 
\and School of Particles and Accelerators, Institute for Research in Fundamental Sciences (IPM), P.O.Box 19395-5531, Tehran, Iran 
\and Department of Physics, Faculty of Nano and Bio Science and Technology, Persian Gulf University, 75169 Bushehr, Iran. 
\and AGH University, Faculty of Physics and Applied Computer Science, Al. Mickiewicza 30, 30-055 Krak\'{o}w, Poland. 
\and Department of Physics, University of Science and Technology of Mazandaran, P.O.Box 48518-78195, Behshahr, Iran.}
\date{Received: April 25, 2024 / Revised version: \today}
%

\abstract{
This article introduces {\tt AAK24}, a Next-to-Leading Order (NLO) QCD analysis of polarized data from 
both polarized Deep Inelastic Scattering (DIS) and Semi-Inclusive Deep Inelastic Scattering (SIDIS) 
experiments on the nucleon. The {\tt AAK24} QCD analysis incorporates SU(2) and SU(3) symmetry breaking, 
specifically $\delta \bar{u} \neq \delta \bar{d} \neq \delta \bar{s}$, while assuming $\delta \bar{s}$ and $\delta s$ are equal. 
Emphasizing the significance of the semi-inclusive data, the study explores the determination of 
polarized sea quark distributions. Recent experimental data from {\tt JLAB17}, {\tt COMPASS16}, 
and {\tt COMPASS17}, including the $A_2$ asymmetry measurements along with SIDIS observables, are 
thoroughly examined for their impact on the central values of polarized PDFs, their uncertainties, 
and overall fit quality.  Additionally, we include the nonperturbative target mass corrections (TMC) as 
well as higher-twist terms (HT) which are particularly important.    In this work, the uncertainties are 
quantified using the standard Hessian method. The main results and findings of the {\tt AAK24} QCD 
analysis show overall good agreement with the analyzed experimental data, aligning well with other 
polarized PDF determinations, particularly {\tt DSSV14}, {\tt LSS10}, {\tt JAM17}, and {\tt AKS14}, 
all considering SU(2) and SU(3) symmetry breaking.
\PACS{
      {PACS-key}{discribing text of that key}   \and   
      {PACS-key}{discribing text of that key}
     } 
} 

\maketitle

\section{Introduction}\label{Introduction}

{ 
	
	In recent years, advancements in the determination of the nucleon  
	partonic distributions and spin projections, 
	particularly from high-energy experimental data, have significantly enhanced our 
	understanding in this area~\cite{Deur:2018roz,Aidala:2012mv,Ethier:2020way,AbdulKhalek:2021gbh,Anderle:2021dpv,COMPASS:2015mhb}. 
	Extracting the polarized and unpolarized partonic distributions from QCD analysis is pivotal for 
	studying the phenomenology of hard scattering processes~\cite{Aidala:2012mv}.	
	The determination of polarized parton distribution functions 
	(polarized PDFs) and along their uncertainties 
	has rapidly expanded, driven by polarized deep inelastic scattering (DIS) 
	experiments at CERN, SLAC, DESY, 
	and JLAB~\cite{DeFlorian:2019xxt,Khan:2022vot,Adamiak:2021ppq,Khanpour:2017cha,Arbabifar:2023hok,Taghavi-Shahri:2016idw,Hirai:2008aj,Blumlein:2010rn,Sato:2016tuz,Leader:2014uua,deFlorian:2009vb}, along with the inclusion of semi-inclusive deep 
	inelastic scattering (SIDIS) data by various 
	phenomenological groups~\cite{Adamiak:2023yhz,deFlorian:2014yva,Arbabifar:2013tma,Leader:2010rb}.
	
	In this work, we utilize comprehensive proton and deuteron SIDIS 
	asymmetry data sets from SMC   and 
	COMPASS group at CERN, including the semi-inclusive asymmetries for charged pion ($\pi^\pm$),  
	kaon ($k^\pm$) and light charged hadron ($h^\pm$) production from a polarized proton target. 
	Using the SIDIS asymmetry data sets, we take into account the flavor 
	SU(2) and SU(3) symmetry breaking as well.
	The recent experimental data from {\tt JLAB17}~\cite{CLAS:2017qga}, {\tt COMPASS16}~\cite{COMPASS:2016jwv}, 
	and {\tt COMPASS17}~\cite{COMPASS:2015mhb} experiments have 
	been incorporated into our data samples. 
	More importantly, this analysis marks the first simultaneous incorporation of SIDIS data and the 
	$A_2$ asymmetry measurements for proton, neutron and deuteron from 
	{\tt SLAC/E143}, {\tt SMC}, {\tt HERMES}, and {\tt Jlab-Hall A} experiments as well. 
	We have thoroughly examined the impact of these new data sets on 
	both the central values and extracted uncertainties, as well as their effect on the 
	overall fit quality. 
	Additionally, we calculate the uncertainty of polarized PDFs 
	using the standard Hessian method.
	
	This paper is organized as follows: Sec.~\ref{Theoretical_Framework} outlines 
	the basic theoretical formalism utilized in the analysis of polarized PDFs within the 
	context of {\tt AAK24}. 
	Sec.~\ref{datasets} discusses the details of the polarized DIS 
	and SIDIS datasets for proton, neutron, and deuteron targets.
	In Sec.~\ref{parametrization}, we present and discuss the {\tt AAK24} parametrization 
	of polarized PDFs and the assumptions that have considered.
	Sec.~\ref{Minimization-strategy} introduces the {\tt AAK24} methodology employed for 
	extracting the polarized PDFs, the fitting procedure, and estimating their uncertainties, 
	specifically through the standard Hessian method.
	Sec.~\ref{Results} presents the full results and findings of this work, 
	comparisons with other models, and discussion of data/theory agreement as well. 
	We also discuss the effects arising from the inclusion of the new data sets added to 
	this analysis. 
	Finally, Sec.~\ref{Conclusions} provides a comprehensive summary and conclusion of the work.
	
}

%
%
\section{Theoretical framework} \label{Theoretical_Framework}

{ 
	
In the following, we present a brief review of the basic
theoretical formalism employed in the analysis of polarized PDFs within the context of {\tt AAK24}.
Our discussion in this section primarily focuses on two key aspects: the spin-dependent
structure function within the framework of QCD, and the unpolarized structure function as well.
The inclusion of the unpolarized structure function is
essential as it plays a pivotal role in extracting the polarized PDFs from the measured asymmetries.
Furthermore, we also present and discuss the observables used in this
work to extract the {\tt AAK24} polarized PDFs.

In perturbative QCD, the spin structure 
function \( g_1 \) for \( Q^2 \gg \Lambda^2 \) (without specifying 
the nucleon target label \( N \)) can be expressed as:

\begin{equation}
g_1(x, Q^2) = g_1^{\rm \tau2 + TMCs}(x, Q^2) + g_1(x, Q^2)_{\mathrm {HT}}~,
\label{g1QCD}
\end{equation}

In this equation, \( g_1^{\rm \tau2 + TMCs}(x, Q^2) \) represents the leading 
twist (\(\tau = 2\)) contribution to the \( g_1 \) structure function, including the 
effects of target mass corrections (TMCs). The term \( g_1(x, Q^2)_{\mathrm{HT}} \) 
accounts for the contributions from higher-twist QCD operators as well.  


The \( g_1(x, Q^2)^{\tau2} \) is the well-known (logarithmic in \( Q^2 \)) NLO perturbative QCD contribution, and 
can be written as,   

\begin{eqnarray}
	\label{g1-ddis}
	g_{1}(x, Q^{2})^{\tau2} &=& \frac{1}{2} 
	\sum_{q, \bar{q}}^{n_f} e_{q}^{2}
	\left\{ \left[1 + 
	\frac{\alpha_{s}}{2\pi}
	\delta {C}_{q}\right] \right.
	\otimes \delta q(x, Q^{2}) \nonumber\\
	&&+ \left. \frac{\alpha_{s}} 
	{2\pi}\: 2\delta
	C_{g} \otimes 
	\delta g(x, Q^{2}) \right\}\,.
\end{eqnarray}

In equation above, \( \alpha_{s} \) represents the strong coupling constant at NLO accuracy, 
and \( \delta q \), \( \delta \bar{q} \), and \( \delta g \) correspond to the polarized quark, antiquark, and gluon density functions, respectively. 
In Eq.~\ref{g1-ddis}, the symbol \(\otimes\) represents the convolution integral. 
The terms \(\delta C_{q}\) and \(\delta C_{g}\) denote the corresponding hard scattering Wilson coefficients, 
which are calculable in perturbative QCD and can be found, for example, in Refs.~\cite{Lampe:1998eu,Arbabifar:2013tma}. 
The explicit twist-2 expression for the \( g_1 \) in the presence of TMCs can be written as~\cite{Dong:2006jm,Dong:2008zg,Dong:2007iv,Dong:2007zzc,Sidorov:2006fi}:

\begin{eqnarray}
\label{eq:g1TMC}
&& g_1^{\tau2 + \mathrm {TMCs}}(x,Q^{2})  \nonumber  \\
& = & \frac{xg_1^{\tau 2}(\xi, Q^{2}; {\mathrm M} = 0)}
{\xi(1 + 4 {\mathrm M}^{2} x^{2}/Q^{2})^{3/2}}   
+  \frac{4{\mathrm M}^{2} x^{2}}{Q^{2}}\frac{(x + \xi)}
{\xi(1 + 4 {\mathrm M}^2x^2/Q^2)^2}\int_{\xi}^{1}\frac{d\xi'}{\xi'} 
g_1^{\tau 2}(\xi', Q^2; {\mathrm M}=0)  \nonumber  \\
& - & \frac{4 {\mathrm M}^{2} x^{2}}{Q^2}
\frac{(2 - 4 {\mathrm M}^{2} x^{2}/Q^{2})}
{2(1 + 4 {\mathrm M}^2x^2/Q^2)^{5/2}}  
\times \int_{\xi}^1\frac{d \xi'}{\xi'} \int_{\xi'}^{1}
\frac{d\xi''}{\xi''} g_1^{\tau 2} (\xi'', Q^2; {\mathrm M} = 0) \,, 
\end{eqnarray}

in which, ${\mathrm M}$ denotes the nucleon mass. 
The parameter \(\xi\) in the above equation refers to the Nachtmann variable~\cite{Nachtmann:1973mr}, which is defined as:
	
\begin{equation}
\xi = \frac{2 x} 
{1 + \sqrt{1 + 4 {\mathrm M}^{2} x^{2} / Q^{2}}}~.
\end{equation}

The maximum kinematic value of the Nachtmann variable \(\xi\) presented above is less than unity, 
implying that both the polarized and unpolarized target mass-corrected 
leading-twist structure functions do not vanish at \(x = 1\).  
Finally, the dynamical higher-twist effects are given by:
\begin{equation}
	g_1(x, Q^{2})_{\mathrm{HT}} = \frac{h(x, Q^{2})}{Q^{2}} + {\cal O}\left(\frac{\Lambda^4}{Q^4}\right)~\,.
	\label{HTQCD}
\end{equation}
This contribution is fitted to the data along with the fit parameters.   

Given that {\tt AAK24} is limited to the NLO accuracy,
consistent with most global polarized PDF analyses available in literature,
we utilize the hard scattering coefficients computed to the same NLO order.
In addition, \( n_f \) is the number of active flavors (\( n_f = 3 \) in our analysis). 
	
In the leading-twist approximation and at the NLO accuracy, the $g_{2}(x, Q^{2})$
spin-dependent structure-function
which contains the transverse spin carried by the
nucleons is given in
terms of the twist-2 component of $g_1(x, Q^{2})$ via
the Wandzura-Wilczek relation, as detailed in Refs.~\cite{Piccione:1997zh}.
The $g_{2}(x, Q^{2})$ can be written as,
	
\begin{equation}
\label{g2-ddis}
g_{2}({x}, Q^{2}) = - 
g_{1}({x}, Q^{2}) + 
\int_{x}^{1}\frac{dy}{y} 
\, g_{1}({y}, Q^{2})\,.
\end{equation}
	
It is important to note that these leading-twist approximation
results are, strictly speaking,
valid in the Bjorken limit $Q^2 \to \infty $ for a given fixed value of $x$.
At the finite values of photon virtuaity $Q^{2}$ the
power-suppressed corrections to the spin-dependent structure function
can make important contributions, particularly in specific kinematic regions.
The simplest of these are the target mass corrections (TMCs) in which
studied in details in our prior analysis~\cite{Khanpour:2017cha},
and it has also been examined in other studies found in the
literature, such as~\cite{Sato:2016tuz,Courtoy:2022kca,Mirjalili:2022cal}.
In the following analysis, we consider the TMC and HT corrections 
 as presented in Eqs.~\ref{eq:g1TMC}  and \ref{HTQCD}.
	
The unpolarized structure function $F_1(x,Q^2)$, which is
essential for extracting the polarized PDFs from the measured
asymmetries, can be expressed as follows:
	
\begin{equation}
\label{F1-sf}
F_{1}({x}, Q^{2})= 
\frac{({1} + \gamma^{2})}
{2 x (1 + R({x}, Q^{2}))} 
F_{2}({x}, Q^{2})\,,
\end{equation}
	
in which is determined per kinetic variables $x$, $R$ and the
nucleon unpolarized structure function $F_{2} (x, Q^{2})$.
In Eq.~\ref{F1-sf}, the factor $\gamma^{2}$ defined as $\frac{4 M^{2} x^{2}}{Q^{2}}$,
and $R$ is the ratio of the longitudinal $\sigma_L$ to the
transverse $\sigma_T$ virtual photoproduction
cross sections; $R = \frac{\sigma_L}{\sigma_T}$~\cite{Whitlow:1990gk}, which in our calculations, we use the experimental values 
measurement by NMC collaboration~\cite{NewMuon:1995zeb}.
	
	The virtual photoproduction longitudinal and transverse asymmetries in the
	Bjorken limit can be expressed as ratios of the
	spin-dependent ($g_{1}({x}, Q^{2})$ and $g_{2}({x}, Q^{2})$) and
	unpolarized ($F_{1}({x}, Q^{2})$ and $F_{2}({x}, Q^{2})$) structure functions.
	
	Finally, having at hand the spin-dependent and unpolarized structure functions,
	one  can define the $A_{1}$ and $A_{2}$ asymmetries for
	polarized DIS processes as follows:

	\begin{equation}
		\label{A1-asy}
		A_{1}({x}, Q^{2}) = \frac{g_{1} ({x}, 
			Q^{2})}{F_{1}({x}, Q^{2})}({1} + 
		\gamma^{2})\,,
	\end{equation}
	
	for $A_1$, and
	
	\begin{equation}
		\label{A2-asy}
		A_{2}({x}, Q^{2})= \gamma \frac{g_{1}
			({x}, Q^{2}) + 
			g_{2}(x,Q^2)}{F_{1}({x}, Q^{2})}\,,
	\end{equation}

	for the  $A_2$ asymmetries, respectively.
	
	We should stress here that, a very large data sample of the $A_{2}$
	asymmetry has been incorporated into the {\tt AAK24} QCD analysis.
	We will provide a comprehensive discussion of this dataset in the forthcoming section.
	The effect of the  $A_2$ asymmetry
	data on the extracted polarized PDFs are studied
	and clearly discussed in Sec.~\ref{Results} as well.
	
	As previously emphasized, one of the primary objectives of this
	study is to incorporate the SIDIS data alongside
	the polarized DIS data, thereby enabling the
	examination of light sea-quark decomposition and to provide well-constraint gluon PDFs.
	In order to increase the statistics of the SIDIS observables,
	we utilize the most recent and precise experimental
	data measured by {\tt JLAB17}, {\tt COMPASS16}, and {\tt COMPASS17} experiments.
	
	At the leading-twist approximation and at NLO accuracy, the spin-dependent
	structure function $g_{1N}^h({x}, {z}, Q^{2})$ for the SIDIS processes can be
	calculated in terms of the polarized PDFs and the 
	Fragmentation Function (FFs). It reads,
	
	\begin{eqnarray}
		g_{1N}^{h,{twist-2}}({x}, {z}, Q^{2})&=&
		\frac{1}{2}\sum_{q, \bar{q}}
		^{n_f}e_{q}^2\left\{\hspace{-0.4cm}
		\phantom{\int\limits_a^b}\left[\delta
		q \left({1} + \otimes \frac{\alpha_s(Q^{2})}{2\pi}
		\delta C_{qq}\otimes\right) {D}_{q}^{h}\right.\right. \nonumber\\
		&&+ \delta q\otimes\frac{\alpha_s(Q^{2})}{{2} \pi}
		\delta {C}_{gq}^{(1)}\otimes {D}_{g}^{h} \nonumber\\
		&&+ \left.\left.\hspace{-0.4cm}
		\phantom{\int} \delta
		g \otimes\frac{\alpha_s(Q^{2})}{{2} \pi} 
		\delta C_{qg}^{(1)}\otimes
		{D}_{q}^{h} \right]({x}, {z}, Q^{2})
		\phantom{\int\limits_{a}^{b}}
		\hspace{-0.4cm}\right\}~, \nonumber\\
		\label{g1h}\,,
	\end{eqnarray}

	in which the $\delta q$, $\delta \bar{q}$ and $\delta g$
	are the polarized quark, antiquark and gluon density functions, the $C_{qq}$, $C_{qg}$ and $C_{gg}$   denote the
	corresponding hard scattering Wilson coefficients. 
	The $D^{h}_{q}$ and $D^{h}_{g}$ are the corresponding FFs of a given hadron.
	
	The unpolarized structure function of $F_{1N}^{h}({x}, {z}, Q^{2})$ for a SIDIS process
	is given by,
	
	\begin{eqnarray}
		\label{F1h-SIDIS}
	F_{1N}^{h,{twist-2}}({x}, {z}, Q^{2}) &=& \frac{1}{2}\sum _{q, \bar{q}}
		^{n_{f}} e_{q}^{2}\left\{\hspace{-0.4cm} \phantom {\int\limits_{a}^{b}}\left[
		q \left(1 + \otimes \frac{\alpha_s(Q^{2})} {{2} \pi}
		C_{qq}\otimes\right) D_{q}^{h} \right.\right. \nonumber\\
		&&+  q \otimes \frac{\alpha_s(Q^2)}{2 \pi} {C}_{gq}^{(1)}
		\otimes D_{g}^{h} \nonumber\\
		&&+ \left.\left.\hspace{-0.4cm} \phantom{\int} {g} \otimes
		\frac{\alpha_{s} (Q^{2})}{2 \pi} {C}_{qg}^{(1)}\otimes
		D_{q}^{h} \right]({x}, {z}, Q^{2}) 
		\phantom{\int\limits_{a}^{b}}
		\hspace{-0.4cm}\right\}~\, \nonumber\\
	\end{eqnarray}

where the $q$ and $g$ are the unpolarized PDFs, the $D^h_{q}$ and $D^h_{g}$ are
the quark and gluon FFs of hadron.
Finlay, the  $C_{q(g) g(q)}$ are the hard scattering
Wilson coefficients available in literature~\cite{deFlorian:1997zj,Arbabifar:2013tma}.
In the following analysis, we use the publicly available
unpolarized PDFs of {\tt CT18} group~\cite{Hou:2019efy}  and for the
quark and gluon FFs of hadron we employ the {\tt DSS}~\cite{DSS} results.

The observable for the SIDIS process that we used in {\tt AAK24} QCD
analysis is the $A^h_{1N}$ asymmetry which can be calculated as
ratio of spin-dependent structure function $g_{1N}^h({x}, {z}, Q^{2})$  and $g_{2N}^h({x}, {z}, Q^{2})$  
 to the unpolarized structure function $F_{1N}^h({x}, {z}, Q^{2})$. 
It is given by~\cite{Leader:2010rb},

\begin{equation}
A_{1N}^h(x,z,Q^2)=
\frac{g_{1N}^h(x, z, Q^{2})-
\gamma^2g_{2N}^h(x, z, Q^{2})}
{F_{1N}^h(x, z, Q^{2})}, 
\label{A1h}
\end{equation}

As in the case of polarized DIS, the factor \(\gamma^2\) 
is defined as \(\frac{4 M^2 x^2}{Q^2}\). 
In this work, we do not include HT and TMC corrections for the case of SIDIS observables. 
The HT and TMC corrections are expected to be less significant due to the kinematic 
region and the precision of the currently available SIDIS data sets~\cite{Leader:2010rb}. 

Having defined the polarized
observables $A_1({x}, Q^{2})$, $A_2({x}, Q^{2})$, 
$A_{1N}^h({x}, {z}, Q^{2})$ and
the structure functions necessary for
the {\tt AAK24} QCD-based analysis, our next step in the
upcoming section is to present an overview of the polarized inclusive DIS
and SIDIS datasets that are subject to analysis in this study.

%
%
\section{Polarized inclusive DIS and SIDIS datasets}\label{datasets}

{

	As we discussed earlier, in {\tt AAK24} QCD analysis, we use
	two types of datasets from polarized DIS and SIDIS
	experiments on proton, neutron and deuteron targets
	which mostly come from relevant experiments
	that were done at DESY, SLAC, JLAB and CERN.
	Our objective is to comprehensively incorporate all
	available data from polarized DIS and SIDIS experiments,
	ensuring coverage across a wide range of kinematic
	variables in terms of both $x$ and $Q^2$.
	In the following, we will discuss in details the
	datasets analyzed in our work.

	From a statistical perspective, the wealth of experimental
	datasets used in our analysis are for the inclusive
	polarized DIS experiments for $A_{1}$ asymmetry for lepton
	scattering from  stationary targets with various combinations
	of targets and lepton spin.

	The published experimental datasets for the $A_{1}$ and $\frac{g_1}{F_1}$
	observables are comprehensively presented in Table.~\ref{table-data-A1}.
	This table provides details on the measured observables, their kinematic
	coverage in terms of $x$ and $Q^2$, as well as references
	to the publications where these datasets can be found.
	
	These datasets have been measured and
	published by various experimental collaborations,
	including {\tt COMPASS}~\cite{COMPASS:2015mhb,COMPASS:2016jwv,COMP1},
	{\tt JLAB}~\cite{CLAS:2017qga,JLabn}, {\tt CLAS}~\cite{CLA1pd}, {\tt EMC}~\cite{EMCp}, {\tt SMC}~\cite{SMCpd}, 
	{\tt SLAC}~\cite{E142n,E143pd,E154n,E155p,E155d},  and {\tt HERMES}~\cite{HERMn,HERMpd}. 
	
	It is worth noting that, in {\tt AAK24} QCD analysis, we have analyzed 
	the latest $A_{1}$ asymmetry data from the
	COMPASS experiment at CERN, encompassing measurements on proton, neutron,
	and deuteron targets, {\tt COMPASS16}~\cite{COMPASS:2015mhb} and {\tt COMPASS17}~\cite{COMPASS:2016jwv}.
	For the case of {\tt COMPASS16} in which
	cover the range from 1 to 190 GeV$^2$ in the photon
	virtuality and from $0.0025$ to $0.7$ in the
	Bjorken scaling variable $x$, they improved the statistical precision of $g_1^p$
	by about a factor of two in the newly explored low-$x$ region $x<0.02$.
	
	The {\tt COMPASS17} data were taken at 160 GeV$^2$ beam energy and the results are presented for
	the kinematic range 1 to 100 GeV$^2$ in photon virtuality, and $4\times10^{-3} < x < 0.7$
	in the Bjorken scaling variable,  and $W>4$ GeV$^2$ in the mass of
	the hadronic final state.

	As indicated in Table.~\ref{table-data-A1}, the recently
	incorporated data from the COMPASS experiment are particularly
	valuable as they offer measurements at the lowest accessible
	values of $x$, (approximately $x \sim 4\times 10^{-3}$).
	Consequently, these data play a crucial role in accurately determining
	the polarization information of gluons and sea
	quarks through inclusive polarized DIS data analysis.
	
	In addition to the COMPASS data, we have incorporated the
	recent $A_{1}$ asymmetry data from the JLAB experiment,
	focusing on measurements with the proton target, as detailed
	in Ref.~\cite{CLAS:2017qga}.
	This inclusion further enriches our dataset for analysis.
	
	The CLAS experiment~\cite{CLA1pd}, contributes significantly to
	our dataset, providing nearly 60 percent of the total data points.
	In total, we include 996 data points for
	the $A_1$ and $\frac{g_1}{F_1}$ asymmetry measurements.
	The E155~\cite{E155p,E155d}, HERMES~\cite{HERMn} and
	JLAB~\cite{JLabn} experiment present special cases of $g_1/F_1$.

	%
	%
	\begin{table*}[!ht]
		\hspace{-10mm}
		\centering \footnotesize{
			{\begin{tabular}{ c|c|c|c|c|c|c|c|c |c c c c c c c  }
					\hline
					Experiment & Process & $N_{\rm data}$ & $x_{\rm min}$ & $x_{\rm max}$ & $Q_{\rm min}^{{2}}$~[GeV$^{2}$] & $Q_{\rm max}^{{2}}$~[GeV$^{2}$] & Observable & $\chi^{2}$ & $\chi^{2}$ (Exc. $A_2$) \\ \hline   \hline
					{\tt COMPASS16}~\cite{COMPASS:2015mhb}&  DIS(p) &~44&0.0045&0.58&1.02&82.04&$A_1^p$& 40.188 &39.361\\
					{\tt COMPASS17}~\cite{COMPASS:2016jwv}& DIS(d) &~43&0.0045&0.569&1.03&74.1&$A_1^d$ &34.269  &33.737\\
					{\tt JLAB17}~ \cite{CLAS:2017qga}& DIS(p) &~10&0.15733&0.58355&1.0522&4.1293& $A_1^p$& 36.312 &36.478\\
					{\tt EMC}~\cite{EMCp}  &    DIS(p)   &  ~10 &0.015&0.466&3.5&29.5&$A_1^p$& 4.100  &4.686\\
					{\tt SMC}~\cite{SMCpd}  &    DIS(p)   &  ~12 &0.005&0.48&1.3&58&$A_1^p$ &4.315 &4.526\\
					{\tt SMC}~\cite{SMCpd}  &    DIS(d)   &  ~12 &0.005&0.479&1.3&54.8&$A_1^d$&  21.428  & 21.261\\
					{\tt COMPASS}~\cite{COMP1} & DIS(p) &  ~15&0.0046&0.568&1.1&62.1&$A_1^p$& 11.601  & 11.186\\
					{\tt SLAC/E142}~\cite{E142n}& DIS(n) &  ~~8&0.035&0.466&1.1&5.5&$A_1^n$ &7.404  &4.445\\
					{\tt SLAC/E143}~\cite{E143pd}& DIS(p) &  ~28&0.031&0.749&1.27&9.52&$A_1^p$&21.111   &20.757 \\
					{\tt SLAC/E143}~\cite{E143pd}& DIS(d) &  ~28&0.031&0.749&1.27&9.52&$A_1^d$&52.322  &50.999 \\
					{\tt SLAC/E154}~\cite{E154n}& DIS(n) &  ~11&0.017&0.564&1.2&15&$A_1^n$  &3.805  &3.803\\
					{\tt SLAC/E155}~\cite{E155p}& DIS(p) & ~24&0.015&0.75&1.22&34.72& $\frac{g_1^p}{F_1^p}$ &29.226  &34.812\\
					{\tt SLAC/E155}~\cite{E155d}& DIS(d) & ~24&0.015&0.75&1.22&34.72& $\frac{g_1^d}{F_1^d}$&21.652 &21.450 \\
					{\tt HERMES}~\cite{HERMn} & DIS(n)  & ~~9&0.033&0.464&1.22&5.25&$A_1^n$&3.939   &3.460 \\
					{\tt HERMES}~\cite{HERMn} & DIS(p)  & ~19&0.028&0.66&1.01&7.36&$\frac{g_1^p}{F_1^p}$ & 16.324 &16.092 \\
					{\tt HERMES}~\cite{HERMpd} & DIS(p)  & ~15&0.0264&0.7248&1.12&12.21&$A_1^p$&10.284  & 9.947 \\
					{\tt HERMES}~\cite{HERMpd} & DIS(d)  & ~15&0.0264&0.7248&1.12&12.21&$A_1^d$  &19.408  &19.569\\
					{\tt JLab-Hall A}~\cite{JLabn}& DIS(n)& ~~3&0.33&0.6&2.71&4.38&$\frac{g_1^n}{F_1^n}$ &2.396  &1.840\\
					{\tt CLAS}~\cite{CLA1pd} &  DIS(p) & 151&0.1088&0.5916&1.01&4.96&$A_1^p$&140.949  & 145.795\\
					{\tt CLAS}~\cite{CLA1pd} &  DIS(d) & 482&0.1366&0.57&1.01&4.16&$A_1^d$ &434.995  & 441.439\\   \hline
					\hline
					{\bf \ TOTAL}: &        &   951 &&&&& &916.038
					  &925.652
					 	 \\
					\hline
		\end{tabular}}}
		\caption{The published experimental data sets for $A_{1}^{N}$ and $\frac{g_{1}^{N}}{F_{1}^{N}}$
			asymmetry measurements from different experiments for different
			polarized DIS target.
			The measured observables, their kinematic range in $x$ and $Q^{2}$,
			and the published references are also shown.
			The individual $\chi^{2}$ values per data point for both the analysis 
			with and without $A_2$ asymmetry data are also presented. 
			The details of the kinematical cuts applied
			on the data and the method on $\chi^{2}$ minimization
			are explained in the text.
		}\label{table-data-A1}
	\end{table*}
	%
	%
	
	In Fig.~\ref{fig1}, the kinematic coverage of the $A_{1}^{N}$ asymmetry
	measurement dataset is represented in the $x$ and $Q^{2}$ plane.
	The data points are categorized by different experiments, as illustrated in the
	presentation of the dataset. As one can see, the newly added data from
	COMPASS experiment offer measurements at the lowest accessible
	values of Bjorken variable $x$, ($x \sim 4\times 10^{-3}$).

\begin{figure}
	\begin{center}
		\includegraphics[clip,width=0.55\textwidth]{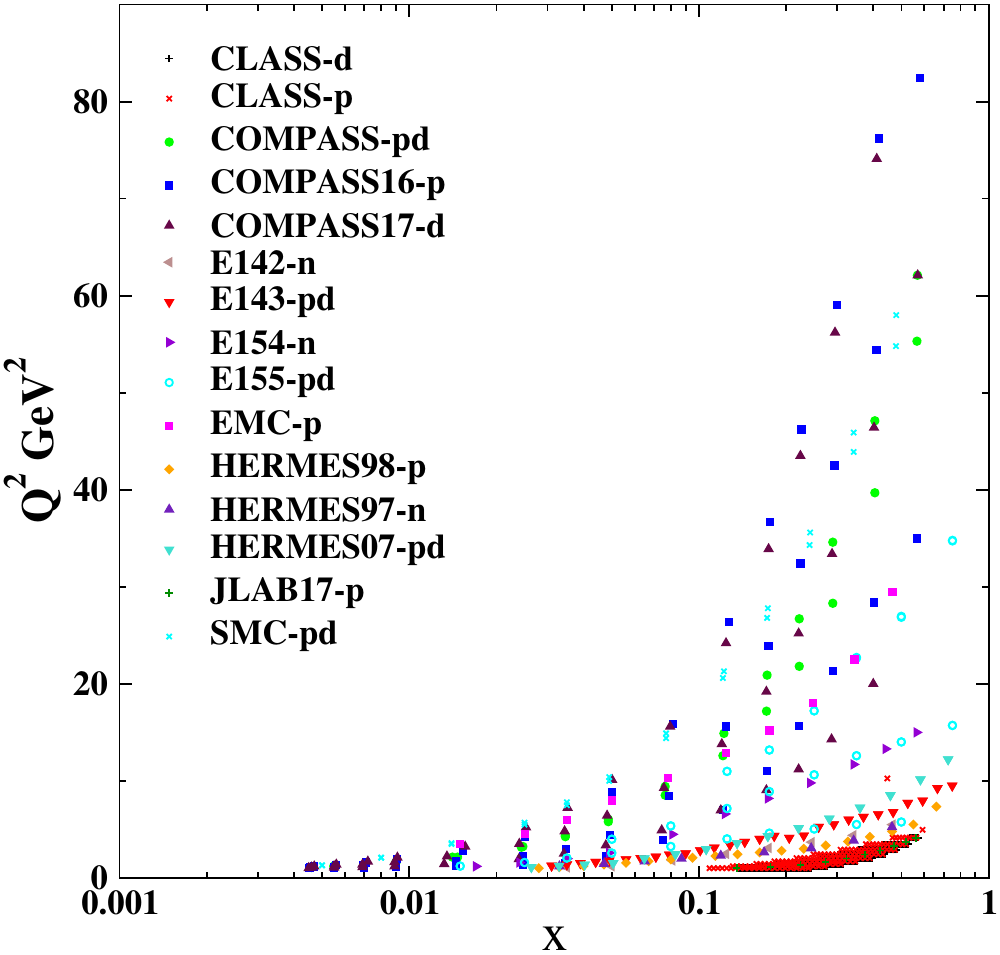}
		\caption{The kinematic coverage in the $x$ and $Q^{2}$ plane of the $A_{1}^{N}$ asymmetry measurement data set.
			The data points are classified by different experiments.}\label{fig1}
\end{center}
\end{figure}

	In the {\tt AAK24} global polarized PDF analysis,
	we incorporate, for the first time, all available worldwide data on the
	inclusive DIS of leptons on proton, neutron, and deuteron targets for
	the measurement of the $A_{2}$ asymmetry measurement.
	This comprehensive inclusion expect to enhance the coverage and
	precision of the analysis.
	
	This inclusion encompasses all datasets from various experiments,
	including {\tt SLAC/E143}~\cite{E143:1995xmc,E143pd,E155:2002iec,E155:1999eug,E154:1997eyc,E154:1997eyc},
	{\tt SMC}~\cite{SpinMuonSMC:1997ixm},
	{\tt JLab-Hall A}~\cite{JLabn}, as well as recent high-precision asymmetry
	measurements from {\tt HERMES}~\cite{HERMES:2011xgd}.
	The datasets for the $A_{2}$ asymmetry measurement are presented in 
	detail in Table.~\ref{table-data-A2}.
	This table provides the kinematic
	coverage in terms of $x$ and $Q^{2}$, as well as references
	to the corresponding publications.
	The $A_2^N$ asymmetry measurements include 140 data points.
    To assess the impact of these data sets, two distinct analyses were 
	conducted in this work, with and without the inclusion 
	of $A_2^N$ asymmetry measurements.   
	A detailed discussion on this matter is provided in Sec.~\ref{Results}.

	%
	%
	\begin{table*}[!ht]
		\hspace{-10mm}
		\centering \footnotesize{
			{\begin{tabular}{ c|c|c|c|c|c|c|c|c c c c c c c c c }
					\hline
					Experiment & Process & $N_{\rm data}$ & $x_{\rm min}$ & $x_{\rm max}$ & $Q_{\rm min}^{2}$~[GeV$^{2}$] & $Q_{\rm max}^{2}$~[GeV$^{2}$] & Observable & $\chi^{2}$ \\ \hline   \hline
					{\tt SLAC/E143}~\cite{E143:1995xmc}& DIS(p) &~12&0.038&0.595&1.49&8.85&$A_2^p$& 7.260 \\
					{\tt SLAC/E143}~\cite{E143:1995xmc}& DIS(d) &~12&0.038&0.594&1.49&8.84&$A_2^d$& 11.441 \\
					{\tt SLAC/E143}~\cite{E143pd}& DIS(p) &  ~12&0.038&0.595&1.49&8.85&$A_2^p$& 7.414 \\
					{\tt SLAC/E143}~\cite{E143pd}& DIS(d) &  ~12&0.038&0.595&1.49&8.86&$A_2^d$& 11.870 \\
					{\tt SLAC/E143}~\cite{E143pd}& DIS(n) &  ~12&0.038&0.595&1.49&8.86&$A_2^n$& 7.414 \\
					{\tt SLAC/E155}~\cite{E155:2002iec}& DIS(p) &~8&0.038&0.78&1.1&8.4&$A_2^p$& 9.130\\
					{\tt SLAC/E155}~\cite{E155:2002iec}& DIS(d) &~8&0.038&0.78&1.1&8.4&$A_2^d$& 14.759\\
					{\tt SLAC/E155}~\cite{E155:1999eug}& DIS(p) &~10&0.022&0.839&1.15&27.18&$A_2^p$& 46.832\\
					{\tt SLAC/E155}~\cite{E155:1999eug}& DIS(d) &~10&0.022&0.839&1.15&27.18&$A_2^d$& 21.102 \\
					{\tt SLAC/E154}~\cite{E154:1997eyc}& DIS(n) &~17&0.017&0.564&1.2&15&$A_2^p$& 0.736\\
					{\tt SMC}~\cite{SpinMuonSMC:1997ixm}& DIS(d) &~4&0.0108&0.228&2.6&18&$A_2^d$& 6.224 \\
					{\tt HERMES}~\cite{HERMES:2011xgd}& DIS(p) &~20&0.039&0.678&1.09&10.35&$A_2^p$&20.394 \\
					{\tt JLab-Hall A}~\cite{JLabn}& DIS(n)& ~~3&0.33&0.6&2.71&4.38&$A_2^n$ &3.369 \\   \hline
					\hline
					{\bf \ TOTAL}: &        &   140 &&&&&& 113.703
					  \\
					\hline
		\end{tabular}}}
		\caption{Same as Table.~\ref{table-data-A1} but this time for the $A_{2}^{N}$ asymmetry measurement.
		}\label{table-data-A2}
	\end{table*}
	%
	%
	
	In Fig.~\ref{fig2}, the kinematic coverage of the $A_{2}$
	asymmetry measurement are presented in the $x$ and $Q^{2}$ plane.
	The data points are categorized by different experiments as well.
	
\begin{figure}
	\begin{center}
		\includegraphics[clip,width=0.65\textwidth]{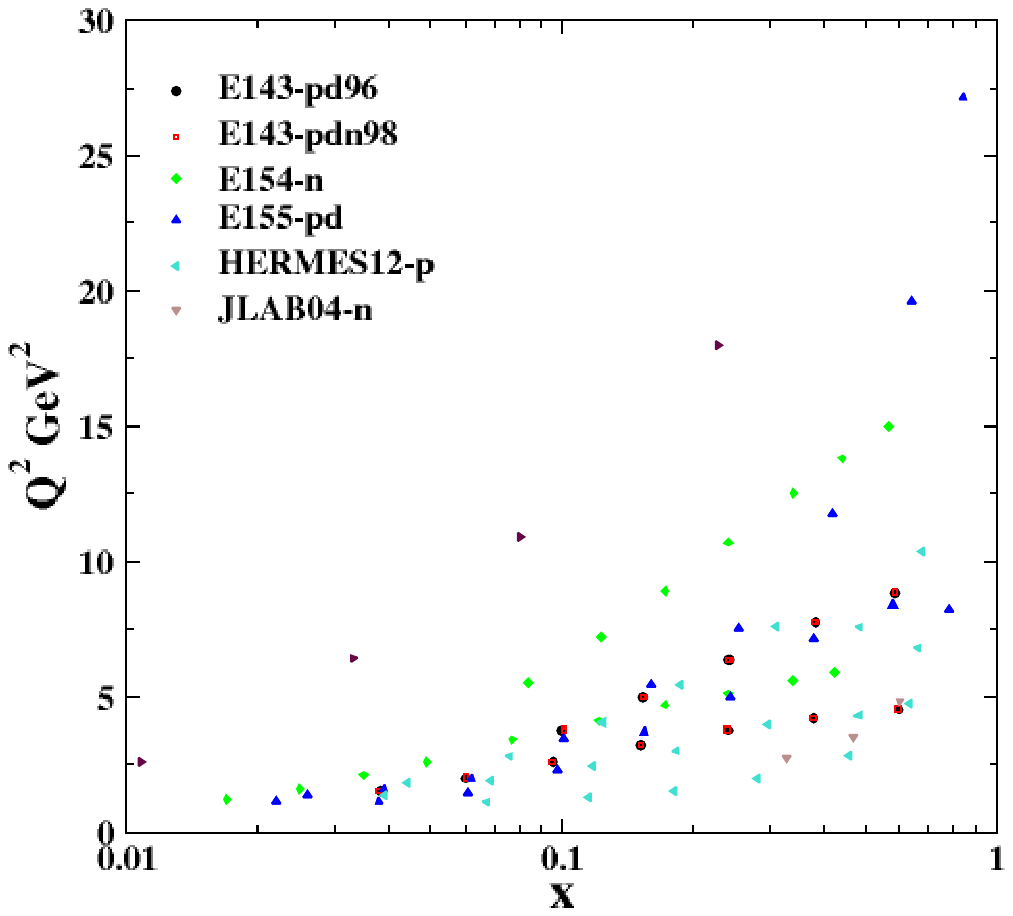}
		\vspace{-4.5cm}
		\caption{The kinematic coverage in the $x$ and $Q^{2}$ plane for
			the $A_{2}^{N}$ asymmetry measurement in which,  for the first time,  
			analyzed in {\tt AAK24} QCD analysis. 
			The data points are classified by different experiments.}\label{fig2}
	\end{center}
\end{figure}

	In our analysis, we have also incorporated all available data for the SIDIS process,
	covering various detected hadrons, including pions ($\pi^{\pm}$),
	kaons ($k^{\pm}$), and light-charged hadrons ($h^{\pm}$).
	This comprehensive addition further enhances the
	richness and diversity of our dataset for analysis.
	The published data points are related to the $A_{1}^{N, h}$ asymmetry, as
 described in Eq.~\ref{A1h},   originating from the SIDIS processes.
	The SIDIS observables are instrumental in enabling the
	consideration of flavor decomposition within the sea-quark density.
	These observables provide valuable information about the different
	flavors of sea quarks within the nucleon.
	
	The measurements of the $A_{1}^{N, h}$ asymmetry,  encompassing data
	from various experiments, incorporate datasets from SMC~\cite{SMC98}, COMPASS08~\cite{COMP1_h}, and COMPASS09~\cite{COMP1_pK}. 
	Furthermore, the recently obtained and highly precise proton data
	from the COMPASS10 experiment~\cite{compass10}, have been included in our QCD analysis.

	The analyzed SIDIS data have been comprehensively summarized in Table.~\ref{table-DIDIS-data}.
	This table provides a classification of the data based on each
	experiment and different detected hadrons, which include pions ($\pi^{\pm}$) and
	kaons ($k^{\pm}$), and light-charged hadrons ($h^{\pm}$).
	In the {\tt AAK24} QCD fit, a   total of 160 data  points from the SIDIS
	process have been utilized.

	%
	%
	\begin{table*}[!ht]
		\hspace{-10mm}
		\centering \footnotesize{
			{\begin{tabular}{ c|c|c|c|c|c|c|c|c |c c c c c c c c }
					\hline
					Experiment & Process & $N_{\rm data}$ & $x_{\rm min}$ & $x_{\rm max}$ & $Q_{\rm min}^{{2}}$~[GeV$^{2}$] & $Q_{\rm max}^{2}$~[GeV$^{2}$] & Observable & $\chi^{2}$ &  $\chi^{2}$ (Exc. $A_2$) \\ \hline   \hline
					{\tt SMC}~\cite{SMC98} & SIDIS(p,$h^{+}$)& ~12&0.005&0.48&10&10&$A_1^{p,h^+}$& 24.973 &26.246\\
					{\tt SMC}~\cite{SMC98} & SIDIS(p,$h^{-}$)& ~12&0.005&0.48&10&10&$A_1^{p,h^-}$&12.096& 12.497\\
					{\tt SMC}~\cite{SMC98} & SIDIS(d,$h^{+}$)& ~12&0.005&0.48&10&10&$A_1^{d,h^+}$ &6.131 &6.514\\
					{\tt SMC}~\cite{SMC98} & SIDIS(d,$h^{-}$)& ~12&0.005&0.48&10&10&$A_1^{d,h^-}$ &16.697&17.004 \\
					{\tt COMPASS08}~\cite{COMP1_h}& SIDIS(d,$h^{+}$)& ~12&0.0052&0.482&1.17&60.2&$A_1^{d,h^+}$&16.108 & 17.906\\
					{\tt COMPASS08}~\cite{COMP1_h}& SIDIS(d,$h^{-}$)& ~12 &0.0052&0.482&1.17&60.2&$A_1^{d,h^-}$&18.210&18.178\\
					{\tt COMPASS09}~\cite{COMP1_pK}& SIDIS(d,$\pi^{+}$)& ~10&0.0052&0.24&1.16&32.8&$A_1^{d,\pi^+}$&11.598& 13.578 \\
					{\tt COMPASS09}~\cite{COMP1_pK}& SIDIS(d,$\pi^{-}$)& ~10&0.0052&0.24&1.16&32.8&$A_1^{d,\pi^-}$&13.476 &12.296 \\
					{\tt COMPASS09}~\cite{COMP1_pK}& SIDIS(d,$k^{+}$)& ~10&0.0052&0.24&1.16&32.8&$A_1^{d,k^+}$&26.377 & 29.621\\
					{\tt COMPASS09}~\cite{COMP1_pK}& SIDIS(d,$k^{-}$)& ~10&0.0052&0.24&1.16&32.8&$A_1^{d,k^-}$&26.354  &29.791 \\
					{\tt COMPASS10}~\cite{compass10}& SIDIS(p,$\pi^{+}$)& ~12&0.0052&0.48&1.16&55.6&$A_1^{p,\pi^+}$ &15.294 &15.853 \\
					{\tt COMPASS10}~\cite{compass10}& SIDIS(p,$\pi^{-}$)& ~12&0.0052&0.48&1.16&55.6&$A_1^{p,\pi^-}$& 13.491&15.303 \\
					{\tt COMPASS10}~\cite{compass10}& SIDIS(p,$k^{+}$)& ~12&0.0052&0.48&1.16&55.6&$A_1^{p,k^+}$&20.450 &22.077 \\
					{\tt COMPASS10}~\cite{compass10}& SIDIS(p,$k^{-}$)& ~12 &0.0052&0.48&1.16&55.6&$A_1^{p,k^-}$&6.673 &7.648\\   \hline
					\hline
					{\bf \ TOTAL}: &        &   160 &&&&& &227.934
					 &244.520
					 			\\
					\hline
		\end{tabular}}}
		\caption{
			The published data points for the $A_{1}^{{N}, {h}}$ asymmetry measured   by SMC and COMPASS collaborations  on
			different targets.
			The measured observables, the number of data points, their kinematic range in $x$ and $Q^{2}$,
			and the published references are also shown.
			The individual $\chi^{2}$ values per data point for both the analysis 
			with and without $A_2$ asymmetry data are also presented. 
		}\label{table-DIDIS-data}
	\end{table*}
	%
	%

	The kinematic coverage of the $A_{1}^{N, h}$
	asymmetry measurement are presented in Fig.~\ref{fig3} in the $x$ and $Q^{2}$ plane.
	The data points are categorized by different experiments as well.

\begin{figure}
	\begin{center}
		\includegraphics[clip,width=0.55\textwidth]{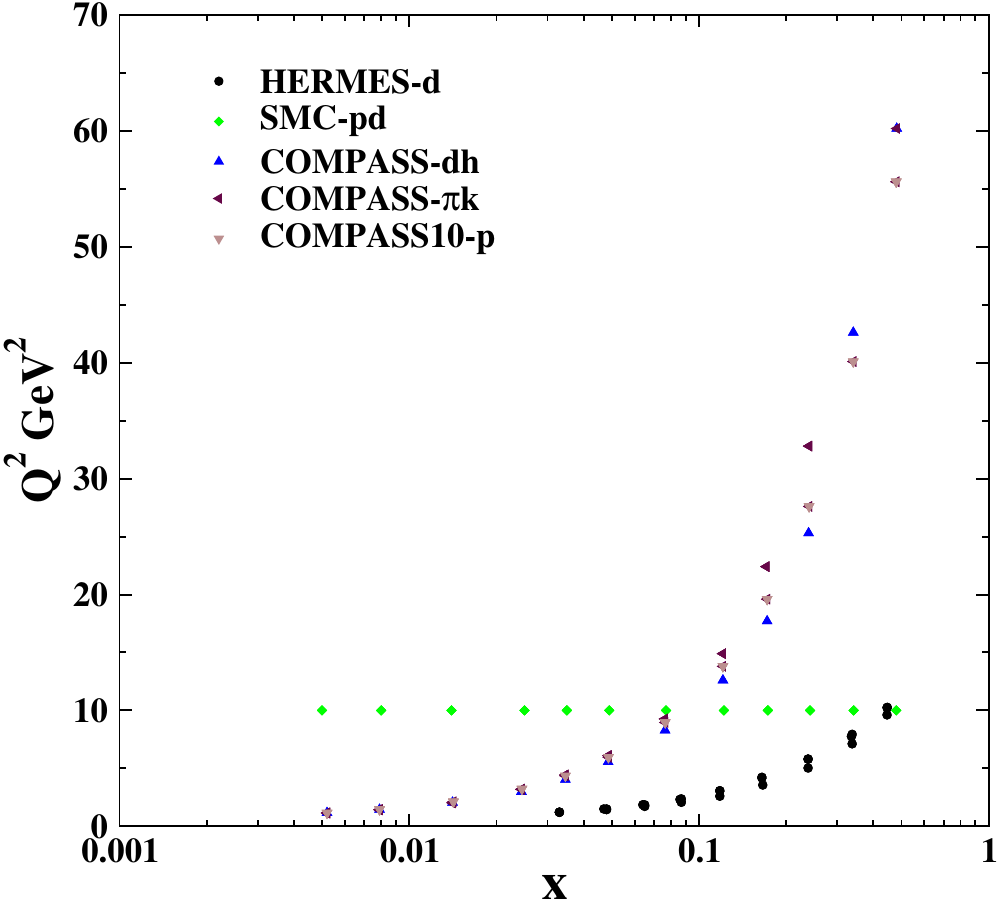}
		\caption{
			The kinematic coverage in the $x$ and $Q^{2}$ plane of the $A_{1}^{N, h}$ asymmetry measurement data set.
			The data points are classified by different experiments.}\label{fig3}
\end{center}
\end{figure}

	All the datasets analyzed in {\tt AAK24} polarized PDFs are subject to the
	cut on the $Q^{2} \geq 1$ GeV$^{2}$, in which
	below $Q^2 = 1$ GeV$^2$ the perturbative QCD is not reliable.
	For all polarized DIS experiments where they are available, we fit
	directly the asymmetries of as $A_{1}$ and $A_{2}$.

%
%
\section{{\tt AAK24} parametrization of polarized PDFs} \label{parametrization}

In this section, we will present and discuss the {\tt AAK24}
parametrization of polarized PDFs and the assumptions that we have considered.
For the generic parametrization of polarized PDFs, assigned to
each parton at a specified input scale of $Q_0^2 = 1 { GeV}^2$,
we choose the following standard functional form,
	
\begin{eqnarray}
\label{eq:parm}
x\:\delta q = {\cal N}_{q} \eta_{q}
x^{a_{q}}({1} - x)^{b_{q}}
(1 + c_{q} x^{0.5} + d_{q} x)\,,
\end{eqnarray}
	
which is defined in terms of five shape fit parameters
$\eta_q$, $a_q$, $b_q$, $c_q$ and $d_q$,
and the normalization factor ${\cal N}_{q}$ as well.
The normalization factor ${\cal N}$ in Eq.~\ref{eq:parm}
is calculated as follows:
	
\begin{eqnarray}
\frac{1}{{\cal N}_q} &=& \left(1 + d_{q}
\frac{a_q}{a_{q} + b_{q} + 1} \right)\,
B\left( a_{q}, b_{q} + 1\right) \nonumber\\
&& + c_{q} B \left(a_q + \frac{1}{2}, b_{q} + 1 \right)~,
\label{eq:norm}
\end{eqnarray}
	
which is chosen such that the $\eta_{q}$ represents the first moments of $\delta
q(x, Q_{0}^{2})$. The $B(a,b)$ indicates to the Euler beta function.
	
Since the {\tt AAK24} QCD analysis considers inclusive
DIS data and the SIDIS observables as well,
we attempt to consider the flavor asymmetric
polarized PDFs and try to fit the PDFs $\delta q = \delta u + \delta\bar{u},
\delta d+\delta \bar{d}, \delta
\bar{u}, \delta\bar{d},  \delta\bar{s}$ and $\delta g$ at the input scale.
	
The observables obtained from the SIDIS processes can, in principle, provide the 
capability to distinguish between the strange quark density ($\delta s$) and its 
antiquark counterpart ($\delta \bar{s}$) through separate production of \( K^+ \) and \( K^- \) mesons. 
However, in practice, our current fit lacks the sensitivity to achieve a clear separation 
between these distributions. Therefore, in line with the polarized PDF analyses conducted 
in Refs.~\cite{Sato:2016tuz,Arbabifar:2013tma}, we make the assumption of a flavor-symmetric 
strange quark distribution, meaning that $\delta{s}({x}, Q^{2}) = \delta{\bar{s}}({x}, Q^{2})$.
To control the behavior of $\delta \bar{u}$ and $\delta \bar{d}$ distributions
over the small $x$ region, additional constraints on the shape parameters
are applied by assuming $a_{\delta u + \delta\bar{u}} = a_{\delta\bar{u}}$, and
$a_{\delta d + \delta\bar{d}} = a_{\delta \bar{d}} = a_{\delta s}$ as well.

In general, polarized PDF analyses utilize two well-known sum rules that relate the first moments of polarized PDFs to the $F$ and $D$ quantities, 
which are evaluated in neutron and hyperon $\beta$-decays \cite{PDG} under the assumption of SU(2) and SU(3) flavor symmetries:

\begin{eqnarray}
a_{3} & = & \Delta \Sigma_u - \Delta \Sigma_d = F + D, \label{a3} \\
a_{8} & = & \Delta \Sigma_u + \Delta \Sigma_d - 2\Delta \Sigma_s = 3F - D. \label{a8}
\end{eqnarray}

\noindent Here, $a_{3}$ and $a_{8}$ represent non-singlet combinations of the first moments of the polarized parton distributions, corresponding to the non-singlet $q_3$ and $q_8$ distributions:

\begin{eqnarray}
q_{3} & = & (\delta u + \delta \overline{u}) - (\delta d + \delta \overline{d}), \\
q_{8} & = & (\delta u + \delta \overline{u}) + (\delta d + \delta \overline{d}) - 2(\delta s + \delta \overline{s}).
\end{eqnarray}

A recent reanalysis of the $F$ and $D$ parameters, with updated $\beta$-decay constants \cite{PDG}, yielded $F = 0.464 \pm 0.008$ and $D = 0.806 \pm 0.008$. We have utilized these updated values in our current analysis. However, since our focus is not on flavor symmetry and we have $\delta \bar{u} \neq \delta \bar{d} \neq \delta s$, we can use the combination of Eqs. \ref{a3} and \ref{a8} as follows:

\begin{eqnarray}
\Delta u + \Delta \bar{u} &=& 0.9275 + \Delta s + \Delta \bar{s}, \nonumber \\
\Delta d + \Delta \bar{d} &=& -0.3415 + \Delta s + \Delta \bar{s}. \label{finalfirstmom}
\end{eqnarray}

We apply these relations in our analysis, thus excluding the parameters that define the first moments of $(\delta u + \delta \bar{u})$ and $(\delta d + \delta \bar{d})$ (i.e., $\eta_{u + \bar{u}}$ and $\eta_{d + \bar{d}}$) from the analysis and obtaining them using Eq.~\ref{finalfirstmom}. The impact of symmetry breaking on the first moments of polarized PDFs has been discussed in detail in the literature \cite{su3,deFlorian:2009vb}.

	It is important to mention that the calculations take into account Isospin
	symmetry for both the proton and neutron.
	This symmetry is defined as follows:
	
	\begin{eqnarray}
		\delta u^{p} = \delta d^{n} \,\,\,; \,  
		\delta \bar{u}^{p} = \delta \bar{d}^{n} \,\,\,;  \,  
		\delta d^{p} = \delta u^{n} \nonumber \\
		\delta \bar{d}^{p} = \delta \bar{u}^{n} \,\,\,;  \,  
		\delta s^{p} = \delta s^{n} \,\,\,;   \,  
		\delta \bar{s}^{p} = \delta \bar{s}^{n} \,.
	\end{eqnarray}
	
	Finally, we should highlight here that
	for some shape
	parameters as presented in Eq.~\ref{eq:parm} may lack strong constraints
	from available data.
	In such cases, additional assumptions become necessary. We will provide a detailed
	discussion of these assumptions, as considered in the {\tt AAK24} analysis,
	in Sec.~\ref{AAK24-polarized-PDFs}.

	%
	%
	\section{Minimization strategy}\label{Minimization-strategy}

	In this section, we will introduce the {\tt AAK24} methodology employed for
	extracting polarized PDFs and for estimating their uncertainties,
	specifically through the standard Hessian method.
	
	The {\tt AAK24} best-fit parameters are determined through the
	minimization of the $\chi^{2}$ function, which is defined as follows:
	
	\begin{eqnarray}\label{chi2}
		\chi_{\rm global}^{2} &=& \sum_{n} {w}_{n} \chi_{n}^{2} ,  \nonumber \\
		\chi_{n}^{2} &=& \left( \frac{{1} - {N}_{n}} {\Delta N_{n}} \right)^{2} + \sum_{i}\left(
		\frac{{N}_{n} A_{1, 2, i}^{\rm data} - A_{1, 2, i}^{\rm theory}}{N_{n} 
			\Delta A_{1, 2, i}^{\rm data}} %
		\right)^2,    \nonumber \\
	\end{eqnarray}

	In the equation above, the quantity $A_{1, 2, i}^{\rm data}$ represents the
	measured value of the observable for the data point $i$ from
	the experimental data set $n$.
	$A_{1, 2, i}^{\rm theory}$ refers to the corresponding theoretical
	value for the $i^{\rm th}$ experimental data point,
	and $\Delta A_{1, 2, i}^{\rm data}$ denotes the experimental uncertainty,
	which is typically the combination of statistical and
	systematic uncertainties computed in quadrature.
	${N}_{n}$ is an overall normalization factor for the data of experimental data set $n$,
	and the ${\Delta{N}_{n}}$ is the experimental normalization uncertainty.
	We allow for a relative normalization shift ${N}_{n}$ between different
	experimental data sets within uncertainties of ${\Delta{N}_{n}}$ quoted
	by the experiments.

	For conducting the fitting process and deriving the {\tt AAK24} best-fit parameters,
	we utilize the publicly available {\tt MINUIT} package,
	which has been developed by CERN~\cite{James:1975dr}.
	A very good  total $\chi^{2}/d.o.f = 1.0528$ value  is
	achieved in these calculations.
	Table.~\ref{table2-fit-parameter} presents the numeric values of
	the {\tt AAK24} fit parameters obtained through the QCD analysis.
	The $\chi^2$ values for the total and individual data sets are presented in
	Tables.~\ref{table-data-A1}, \ref{table-data-A2} and \ref{table-DIDIS-data}
	respectively.
	The detailed discussions on the obtained best fit parameters, and
	the fit quality will be presented in Sec.~\ref{AAK24-polarized-PDFs}.
	
	At this point, we are ready to introduce the Hessian method for
	the estimation of polarized PDF uncertainty.
	The standard Hessian method employed in the {\tt AAK24} QCD analysis was
	originally developed in Refs.~\cite{Martin:2002aw,Pumplin:2001ct}.
	This method essentially aims to approximate the posterior
	distribution as a Gaussian distribution within the parameter space.
	To do this end, one first needs to find the set
	of parameters $\bf{a}_0$ corresponding to
	the maximum a posteriori estimate.
	These the set of parameters could be obtained by minimizing a
	certain the $\chi^{2}$ function in Eq.~\ref{chi2}.
	In a neighborhood of $\bf{a}_0$ parameters, the $\chi^{2}(\bf{a})$ function
	can be expanded up to the quadratic term in its parameters
	and the posterior approximated by a multi-dimensional Gaussian function,
	
	\begin{eqnarray}
		p (\textbf{{a}}|\textbf{{m}})
		\propto \exp \Big(- \frac{{1}} {{2}} 
		\chi^{{2}} (\textbf{a}, {m}) \Big)
		\propto \exp \Big(-\frac{1}{2} \Delta \textbf{a}^{T} \, 
		H \, {\Delta} \textbf{a} \Big)\,. \nonumber \\
	\end{eqnarray}
	
	In the last term, $\Delta \textbf{a} = \textbf{a} - \textbf{a}_{0}$, 
	the Hessian matrix elements are given by
	
	\begin{equation}
		H_{ij}
		= \frac{{1}}{{2}}
		\left.
		\frac{\partial^{{2}} \chi^{{2}} (\textbf{a})}
		{\partial {a}^{{i}} \partial {a}^{j}}
		\right|_{\textbf{a} = \textbf{a}_{0} } \,,
		\quad
		i,j = 1, \dots {{n}}_{\rm par}\,.
		\label{eq:hessmatrix}
	\end{equation}
	
	and $\exp(-\frac{1}{2} {\chi}^{2}(\textbf{a}_{0}))$ has been
	absorbed into the normalization of the posterior.
	
	Additionally, the tolerances can be fine-tuned by setting a criterion that the ${\chi}^{2}$
	must increase above its minimum by a specified $\Delta {\chi}^{2}$ value,
	as discussed in Ref.~\cite{Accardi:2021ysh}.
	In the {\tt AAK24} QCD analysis, fits will be conducted on a statistically
	consistent set of polarized DIS data, and thus, there is no need to 
	employ a tolerance criterion.
	Nevertheless, it is important to calculate the tolerance explicitly in
	order to account for any actual deviations from Gaussian behavior
	and to obtain precise estimates of the variance.
	In the results that will be presented in the following Sec.~\ref{Results},
	we take the tolerance criterion into account by setting $\Delta {\chi}^{2} = T = 1$.

}


%
%
\section{{\tt AAK24} fit results}\label{Results}

{

	This section includes the main results and findings of
	{\tt AAK24} polarized PDFs obtained from
	QCD analysis of polarized DIS and SIDIS data.
	We first present the {\tt AAK24} polarized PDFs, the fit
	parameters, and the results for different Parton species.
	Then we compare the {\tt AAK24} NLO theoretical predictions
	with the analyzed data.
	We also compare our results with other polarized PDFs
	available in the literature by discussing in detail
	the similarity and differences between
	these different polarized PDF sets.

}

%
%
\subsection{AAK24 polarized PDFs} \label{AAK24-polarized-PDFs}

{

	This section presents the {\tt AAK24} polarized PDFs at the input scale,
	the extracted best-fit parameters, and
	assumptions considered for the shape fit parameter as well.
	
	The best fit parameters for {\tt AAK24} are presented in Table.~\ref{table2-fit-parameter}.
	Upon examining the numbers presented in this table, several remarks are pertinent.
	In general, a QCD analysis of polarized DIS experiments should incorporate
	additional constraints on the shape parameters.
	This necessity arises from the affects of available data to fully
	constrain all parameters and the limited kinematic coverage of the data.
	Given that the existing experimental data sets lack the power to comprehensively
	determine the behavior of polarized PDFs,
	certain assumptions must be taken into consideration.

	Firstly, to control the small $x$ behavior of the $x\delta \bar u$, $x\delta \bar d$ and $x\delta s = x\delta \bar s$,
	we assume $a_{\delta u + \delta \bar u} = a_{\delta \bar u}$ and
	$a_{\delta d + \delta \bar d} = a_{\delta \bar d} = a_{\delta \bar s}$.
	Since we have determined that the best-fitted values
	of $b_{\delta \bar u}$, $b_{\delta \bar d}$, and $b_{\delta s = \delta \bar s}$ are not strongly
	constrained by the QCD fit and are very similar, all approximately equal to 10,
	we have chosen to maintain them as fixed values at 10.
	
	Furthermore, we have observed that the parameter $c_q$ for certain parton
	species, such as $x\delta u + x\delta \bar u$, 
	$x\delta d + x\delta \bar d$, $x\delta \bar s$, and $x\delta g$, is
	extremely close to zero. As a result, we have opted to set these parameters to
	a fixed value of 0. Additionally, it is worth noting that some of the $c$ and $d$ shape
	parameters for the $x\delta \bar d$ density have been determined with relatively large errors.
	Incorporating additional high-energy polarized data sets from various types of
	experiments has the potential to alleviate some of the extra assumptions we
	have made regarding the shape parameters of the polarized distribution.
	To be more specific, the inclusion of proton-proton collision data from RHIC can
	have a significant impact on our understanding of the distributions
	within the nucleon, particularly in relation to the gluon PDFs.

	\begin{table*}[ht]
		\centering  \footnotesize{
			\begin{tabular}{ c|c|c|c|c|c}
				\hline
				Flavor &  $\eta_{q}$  &  $a_{q}$ &  $b_{q}$ & $c_{q}$& $d_{q}$ \\ \hline  \hline
				$\delta u+\delta\bar{u}$& 0.863 & 0.4126$ \pm$ 0.0079 & 2.4636$\pm$ 0.0174 &$0.0^*$& $40.9115\pm+1.38$\\
				$\delta d+\delta\bar{d}$& -0.4053 & 0.2487$\pm$ 0.0059 & 4.5203$\pm$ 0.00272 &$0.0^*$&$100.9866\pm10.92$\\
				$\delta \bar{u}$& 0.0639$\pm$0.175 & 0.4126$ \pm$ 0.0079 &10.0$^*$  & 12.5012$\pm$ 0.923&$-39.5083\pm8.09$ \\
				$\delta \bar{d}$&-0.0288$\pm$  0.355 & 0.2487$\pm$ 0.0059& 10.0$^*$ & 	102.436$\pm$52.07&$800.569\pm284.799$\\
				$\delta \bar{s}$ &-0.0433$\pm$ 0.0017 &0.2487$\pm$ 0.0059 &10.0$^*$ 	&$0.0^*$&$-28.8485\pm0.0027$\\
				$\delta g$ &  -0.3788$\pm$0.0026 &  2.9019 $\pm$ 0.0908 & 10.0$^*$ &$0.0^*$ & $-3.271 \pm 0.0875$\\  \hline
				\hline
		\end{tabular}}
		\caption{The best-fitted values and their associated statistical errors are
			provided at the input scale $Q_0^2=1$ GeV$^2$. It is important to note that
			parameters marked with an asterisk (*) have been held fixed. }
		\label{table2-fit-parameter}
	\end{table*}

As mentioned earlier, we also take into account the TMC and HT corrections, which play important roles in certain kinematic regions of 
$x$ and Q$^2$. 
The values of the HT corrections extracted from the data in our analysis are presented in Table.~\ref{tab:HT_values}. 

\begin{table}
\begin{center}
\caption{\label{tab:HT_values} The values of HT corrections
extracted from the data in a model independent way. }
\begin{tabular}{ccccccc}
				\hline
$x_i$ &~~~ $h^p(x_i)~[\mathrm {GeV}^2]$ &~~~~~~ $x_i$&~~~ $h^n(x_i)~[\mathrm {GeV}^2]$ \\ \hline  \hline
0.028 &~~~ 0.03092~$\pm$~0.00817 &~~~~~ 0.028 &~~~ 0.0418~$\pm$~0.0170 \\
0.100 &~~~ -0.08545~$\pm$~0.00386 &~~~~~ 0.100 &~~~ 0.0556~$\pm$~0.0140 \\
0.200 &~~~ -0.04048~$\pm$~0.00270 &~~~~~ 0.300 &~~~ -0.01539~$\pm$~0.0108 \\
0.350 &~~~ -0.03238~$\pm$~0.00274 &~~~~~ 0.300 &~~~ 0.008924~$\pm$~0.00581 \\
0.600 &~~~ -0.07021~$\pm$~0.00443 &~~~~~ 0.500 &~~~ -0.008407~$\pm$~0.00374\\  \hline  \hline 
\end{tabular}
\end{center}
\end{table}

    In Fig.~\ref{fig5-AAK24partonQ0errorbar2},  we display the {\tt AAK24} 
	polarized PDFs at the input scale
	for all parton species including $x\delta u+x\delta \bar{u}$, 
	$x\delta d+ x\delta \bar{d}$, $x\delta \bar{u}$, 
	$x\delta \bar{d}$, $x\delta \bar{s}$ and $x\delta g$. 
	The results for the case where we have excluded the measurements of $A_2^p$, $A_2^d$, and $A_2^n$ 
	asymmetries are also presented for the purpose of comparison.
	The ability of determining the polarized sea-quark distributions
	are due to the impact of SIDIS data in which added to our data sample.
	The error bands, obtained through the Hessian method, are also shown as well.
	Upon examining the distributions in this figure, it becomes evident that the sea
	quark densities $x\delta \bar{u}$ and $x\delta \bar{d}$ exhibit 
	relatively large uncertainties.
	Additionally, the input strange sea-quark
	distribution $x\delta s = x\delta {\bar s}$ displays the typical sign change.

\begin{figure}
\begin{center}
			\includegraphics[clip,width=0.32\textwidth]{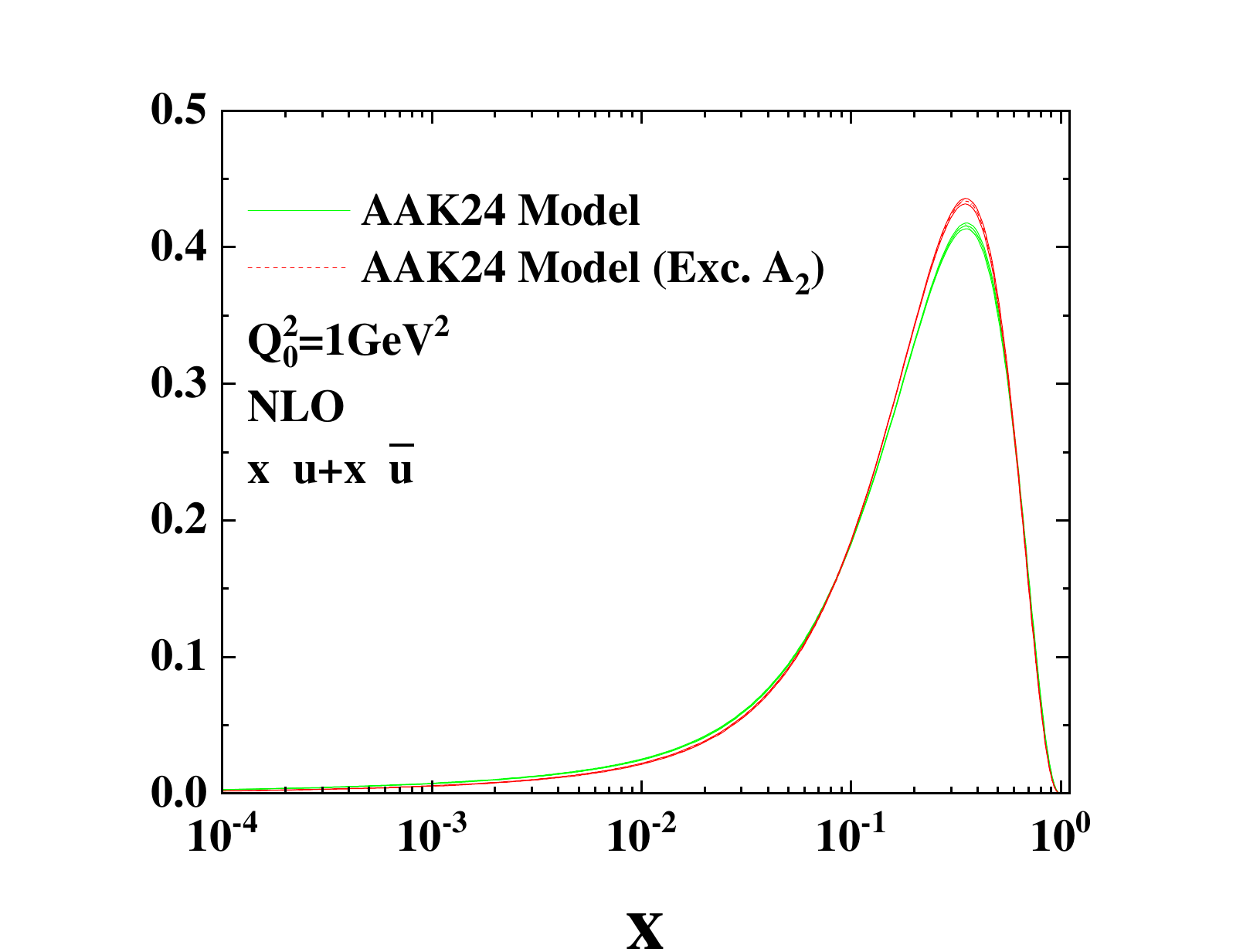}
			\includegraphics[clip,width=0.32\textwidth]{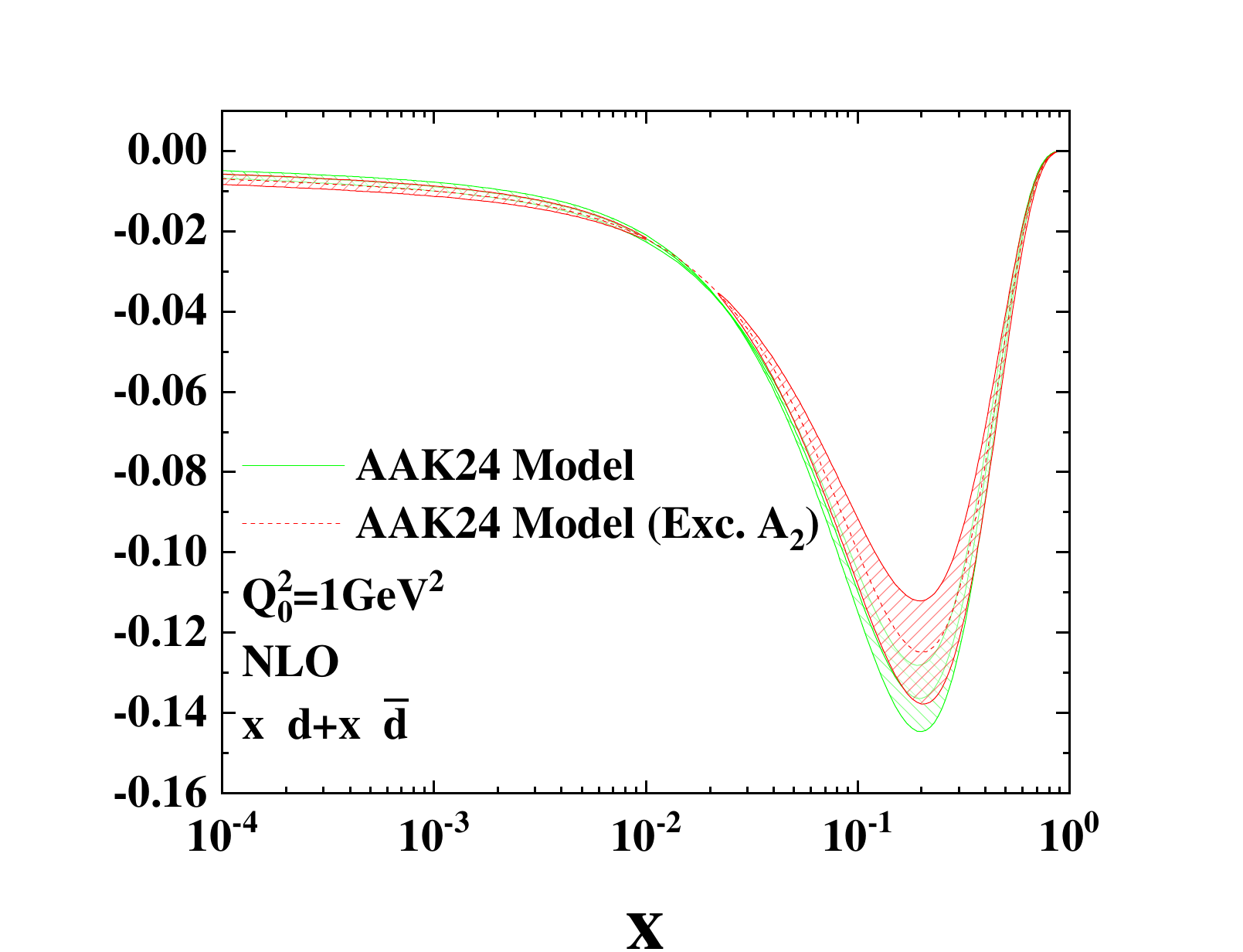}
			\includegraphics[clip,width=0.32\textwidth]{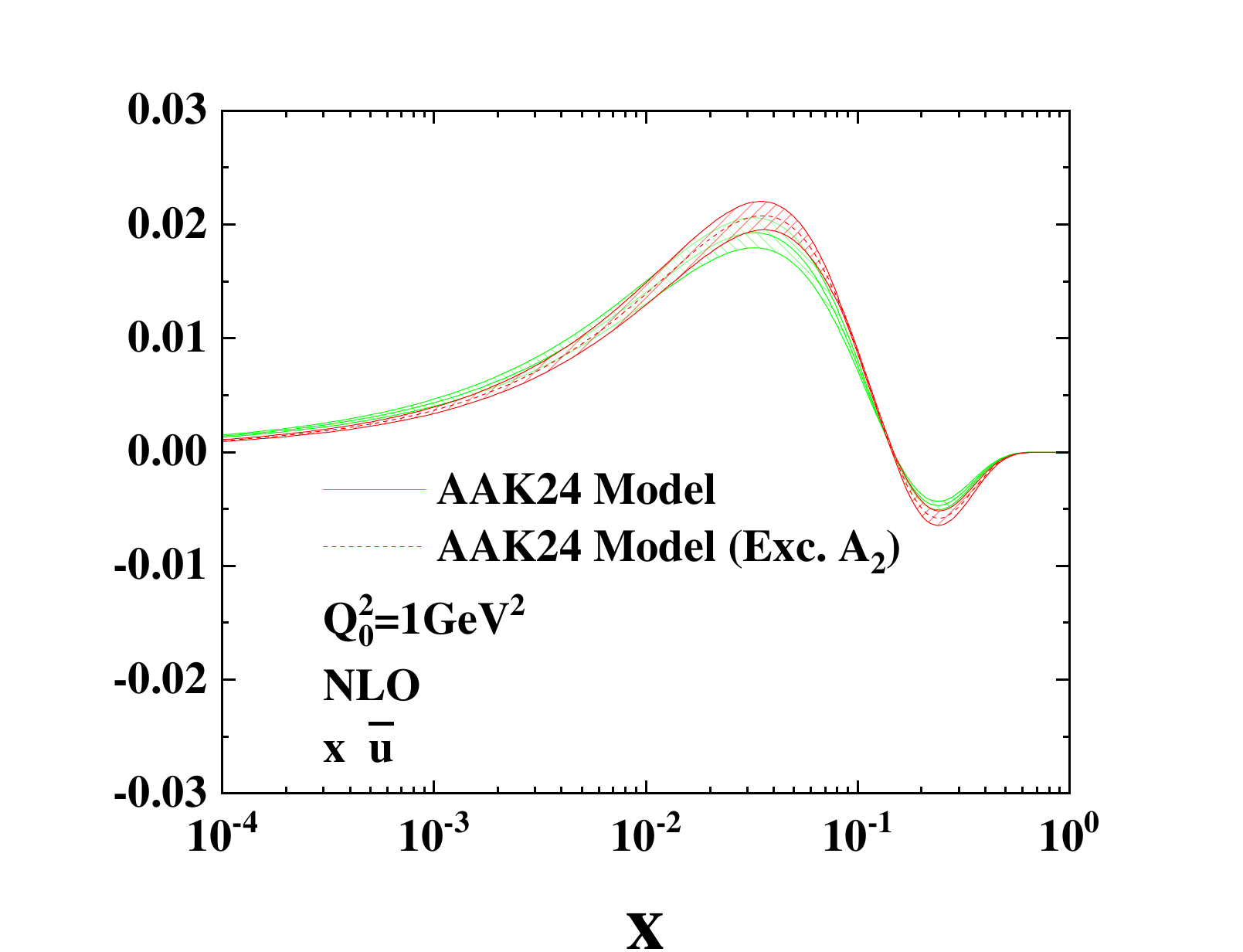}
			\includegraphics[clip,width=0.32\textwidth]{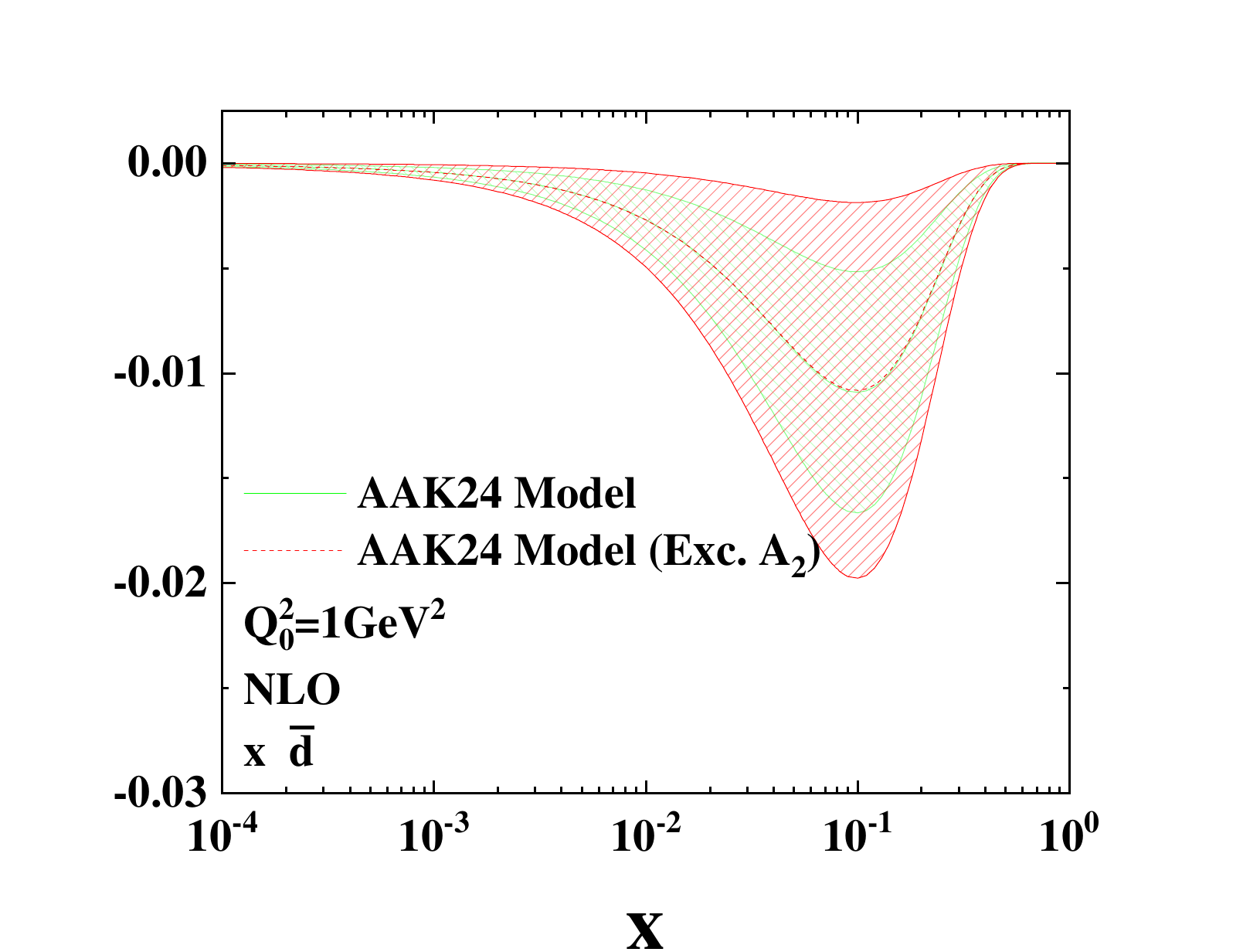}
			\includegraphics[clip,width=0.32\textwidth]{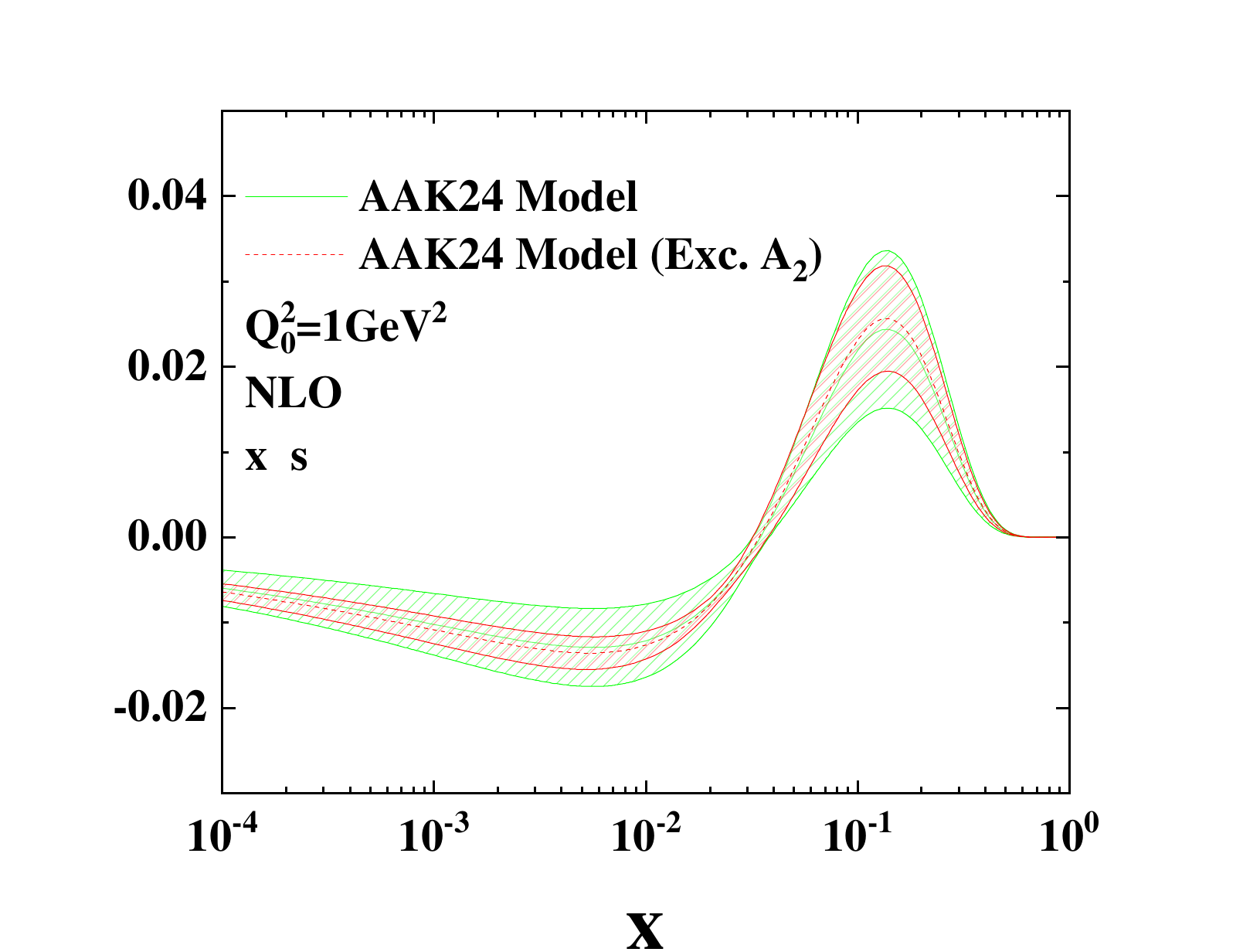}
			\includegraphics[clip,width=0.32\textwidth]{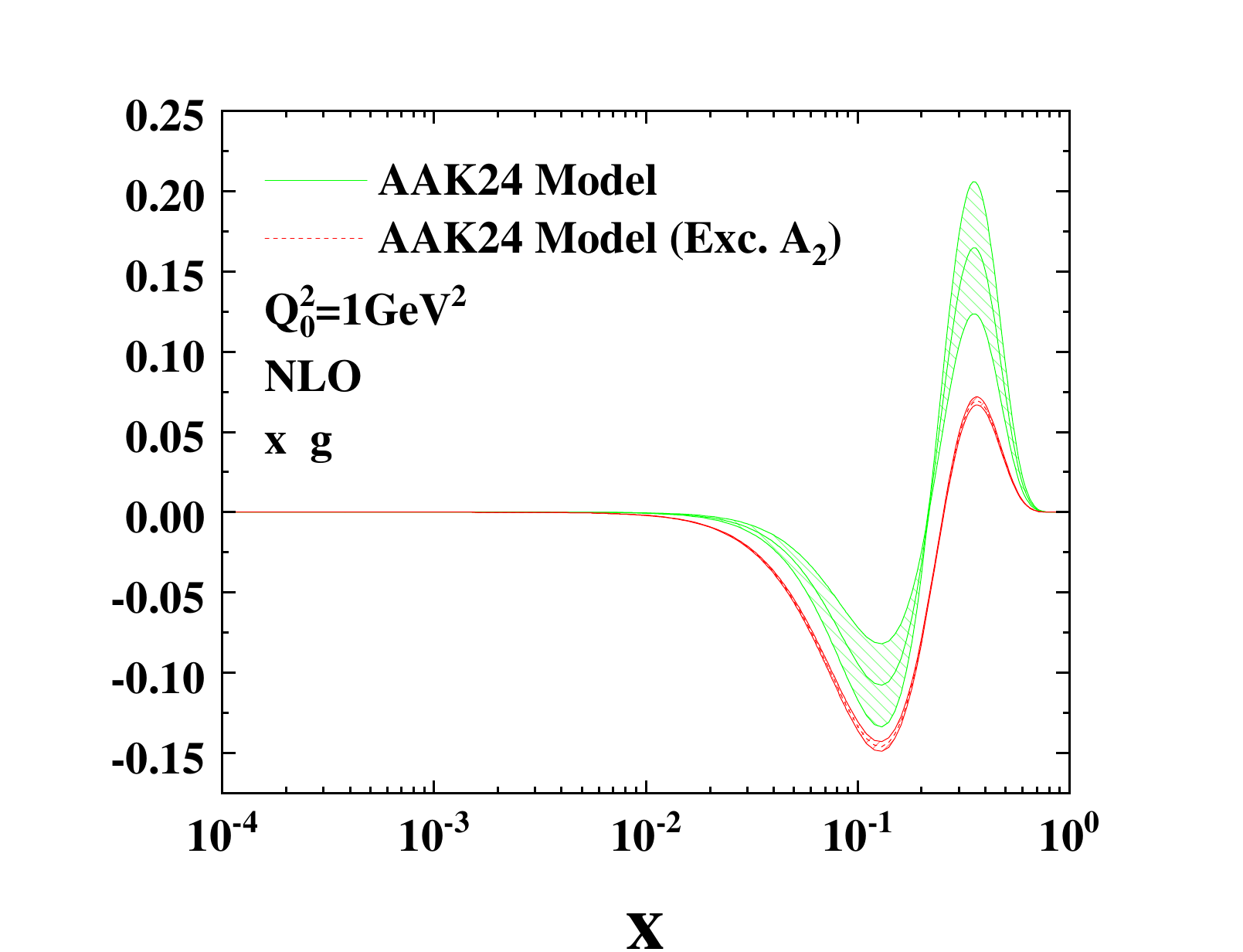}
\caption{The {\tt AAK24} quark and gluon helicity distributions at the input
scale $Q^2_0=1$ GeV$^2$ with and without the $A_2$ asymmetry measurements.
The uncertainty bands, determined using the Hessian method, are also shown as well.}
\label{fig5-AAK24partonQ0errorbar2}
\end{center}
\end{figure}

Let us now discuss the results obtained by excluding the $A_2$ asymmetry measurements to examine 
the effect of these data sets on both the central values and extracted uncertainties.
As observed in Fig.~\ref{fig5-AAK24partonQ0errorbar2}, the inclusion of the $A_2$ asymmetry 
data not only affects the central values of certain parton species but also impacts the error bands. 
These effects are most pronounced in the case of $x\delta \bar{u}$, $x\delta \bar{d}$, $x\delta s$, and gluon densities.	
The most significant changes are noticeable in the case of the gluon PDF in terms of central 
value, and in $x\delta \bar{d}$ and $x\delta s$ in terms of error bands. In terms of individual 
parton flavors, it can be observed that for $x\delta u + x\delta \bar{u}$ and $x\delta d + x\delta \bar{d}$, 
the two results exhibit a similar shape, with a slight reduction of error bands for $x\delta d + x\delta \bar{d}$.
In the case of $x\delta \bar{u}$, when compared to the analysis excluding $A_2$, it is evident that 
the distribution is slightly increased at large values of $x$ and slightly decreased over the medium to small range of $x$, 
with the same size of error bands spanning the entire range of $x$. Regarding the $xs$ density, in general, 
the shape remains nearly unchanged while the error band is increased.
In the case of the gluon PDF, the inclusion of the $A_2$ asymmetry data significantly affects the central value, 
while the error band remains unchanged. Finally, for $x\delta \bar{d}$, the central values are the same, 
but the analysis including the $A_2$ asymmetry measurements leads to smaller error bands.

}

{

	%
	%
	\subsection{Comparison with other polrized PDFs} \label{Comparison-with-other-polrized-PDFs}

	In this section, we present the \texttt{AAK24} polarized PDFs
	and provide detailed comparisons with other results available in the literature,
	namely the results from \texttt{DSSV14}~\cite{deFlorian:2014yva},
	\texttt{AKS14}~\cite{Arbabifar:2013tma}, \texttt{LSS10}~\cite{Leader:2010rb}, 
	and {\tt JAM17}~\cite{Sato:2016tuz}    
	all evaluated at the input scale $Q^2_0 = 1$ GeV$^2$ and at higher value of 10 GeV$^2$.
	The comparisons are displayed in Figs.~\ref{fig4-AAK24parton_1} and \ref{fig4-AAK24parton_2}
	at both the input scale of $Q^2=1$ GeV$^2$ and a higher value of 10 GeV$^2$.

	Examining the $x\delta u+ x\delta \bar{u}$ distribution, 
	generally speaking, one can observe
	that all of the fits are in very good agreement with each other.
	In the case of the $\delta d+ \delta \bar{d}$ distribution, all the
	groups agree very well, except for the \texttt{DSSV14} result, which
	is smaller than those of the others at the input scale. 
	However, the differences are within the uncertainty.  
	As for $Q^2=10$ GeV$^2$,  the \texttt{AAK24} polarized PDFs, {\tt JAM17}  
	and \texttt{LSS10} agree over the entire $x$ region.
	However, for all other parton species, clear differences can be 
	observed among the results of various groups. 
	
	A significant difference can be observed in the case of
	the sea-quark densities $x\delta \bar u$ and $x\delta \bar d$.
	For the case of $x\delta \bar u$, the \texttt{AAK24} results are larger than those of
	other groups for $x<0.1$ at the input scale.
	However, the difference becomes less pronounced at the higher scale.
	Despite the similar results between \texttt{AAK24} and \texttt{JAM17}, 
	for the $x\delta \bar d$ density at both input scale and higher values of Q$^2$ = 10 GeV$^2$, 
	the other results exhibit different behaviors at 

	Examining the strange distribution, we can observe that we
	obtained very similar results compared to those of \texttt{JAM17} at the input scale.
	For the case of the gluon density $x\delta g$,
	the difference between all groups are more pronounced in this case.
Like for the case of \texttt{LSS10}, our gluon density is smaller than those of others 
for medium value of $x$; ($x \sim 0.1$)

	Focusing on the {\tt JAM17}~\cite{Sato:2016tuz} polarized PDFs, 
	one can observe significant differences in their gluon 
	density $x\delta g$,  $x\delta \bar u$,  $x\delta \bar d$, and strange distribution compared to all 
	other distributions presented in Fig.~\ref{fig4-AAK24parton_1}. 
	Distinctions are also more apparent for $x \delta \bar{u}$ and $x \delta \bar{d}$, and $x\delta g$.  
	It is essential to highlight that the {\tt AAK24} and {\tt JAM17} are very 
	similar in the case of $x (\delta u + \delta \bar{u})$ and $x (\delta d + \delta \bar{d})$ and 
	strange distribution.

	These findings hold at a higher scale of 10 GeV$^2$ as well, as depicted in Fig.~\ref{fig4-AAK24parton_2}. 
	Despite the similarity in $x (\delta u + \delta \bar{u})$ and $x (\delta d + \delta \bar{d})$, 
	differences become more pronounced for the gluon density $x\delta g$, sea quarks and 
	strange distribution $x \delta s$. 
	It is crucial to emphasize that the {\tt JAM17} analysis employs the 
	iterative Monte Carlo fitting technique, generating stable fits to 
	polarized PDFs with statistically rigorous uncertainties.

\begin{figure}
\begin{center}
\includegraphics[clip,width=0.32\textwidth]{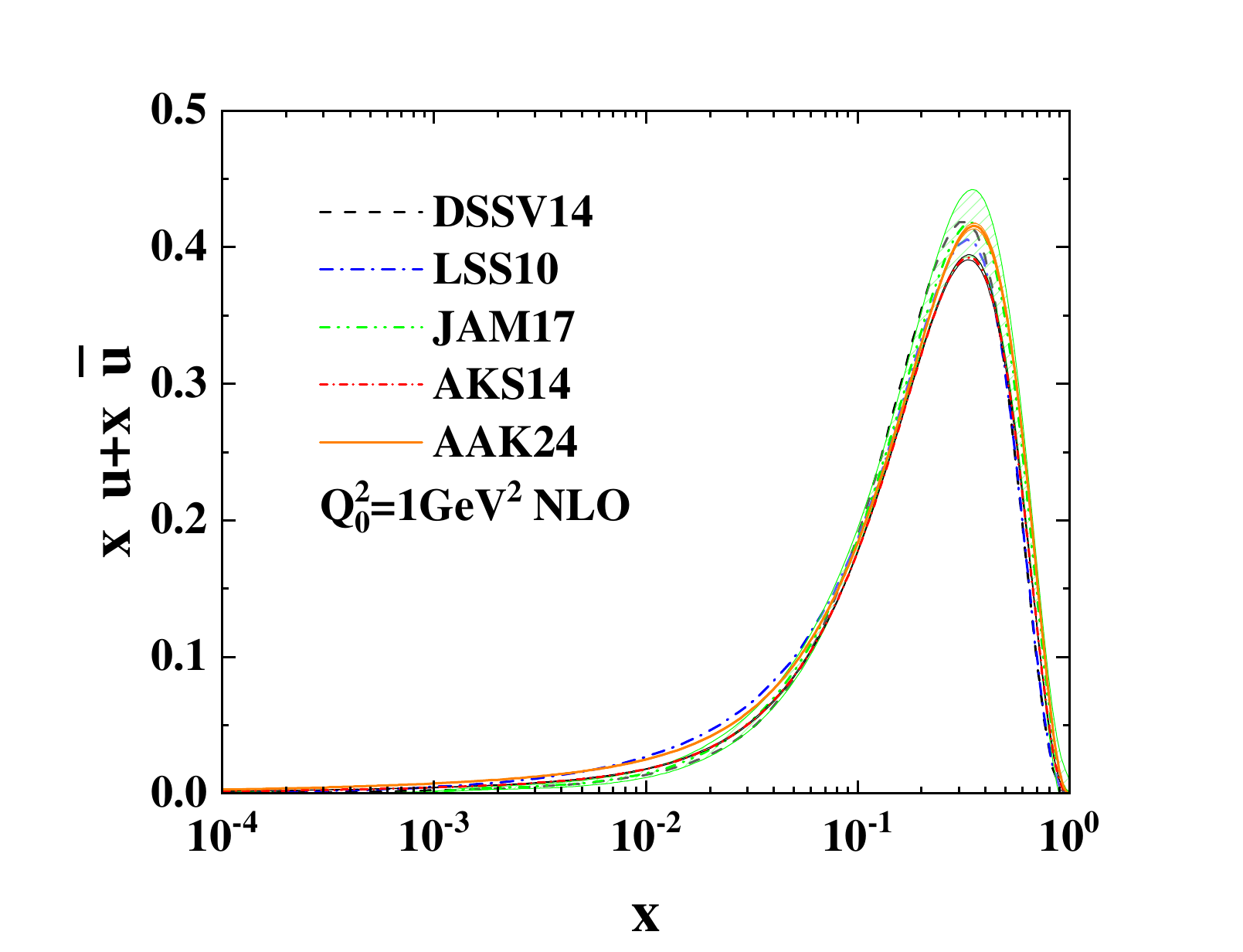}
\includegraphics[clip,width=0.32\textwidth]{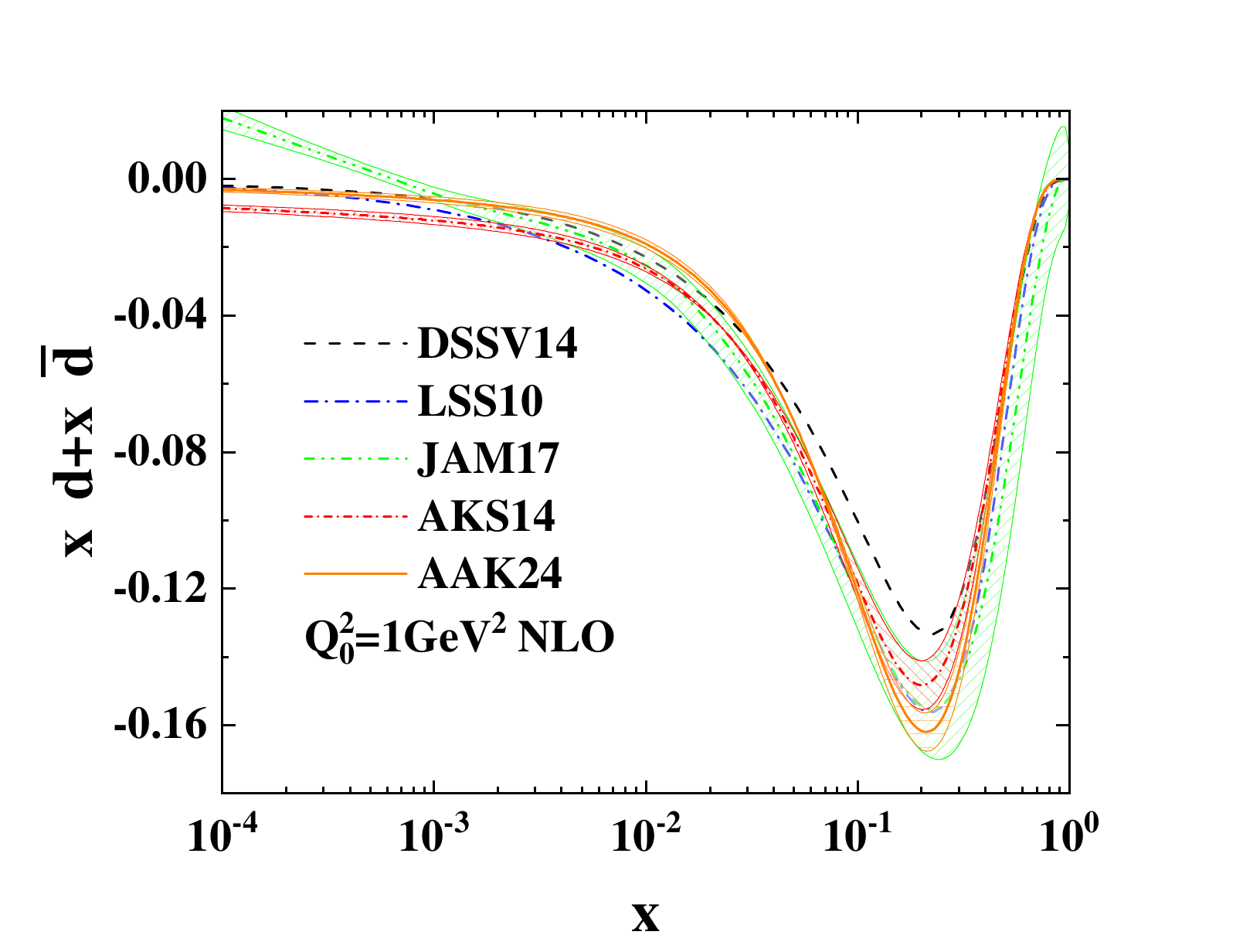}
\includegraphics[clip,width=0.32\textwidth]{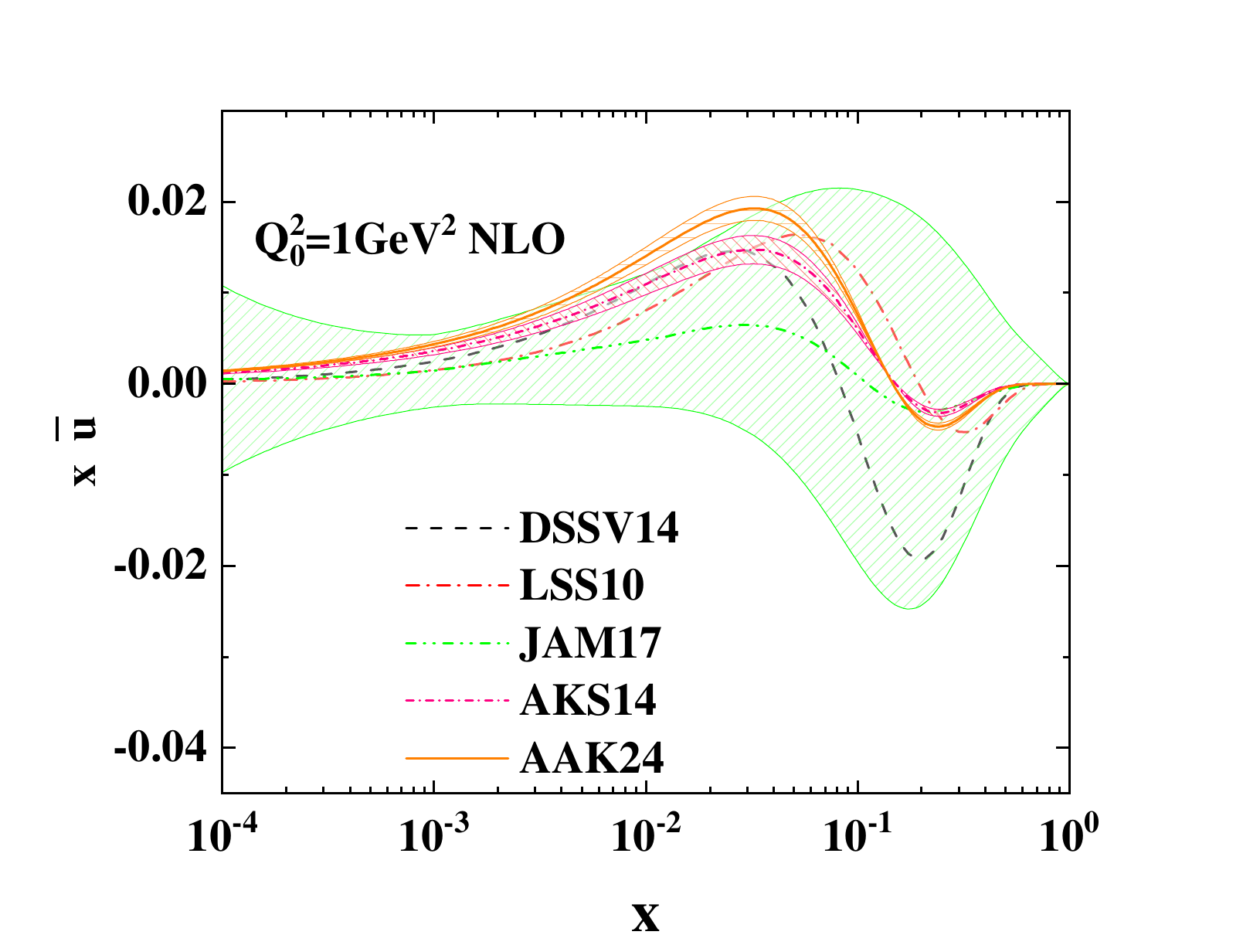}
\includegraphics[clip,width=0.32\textwidth]{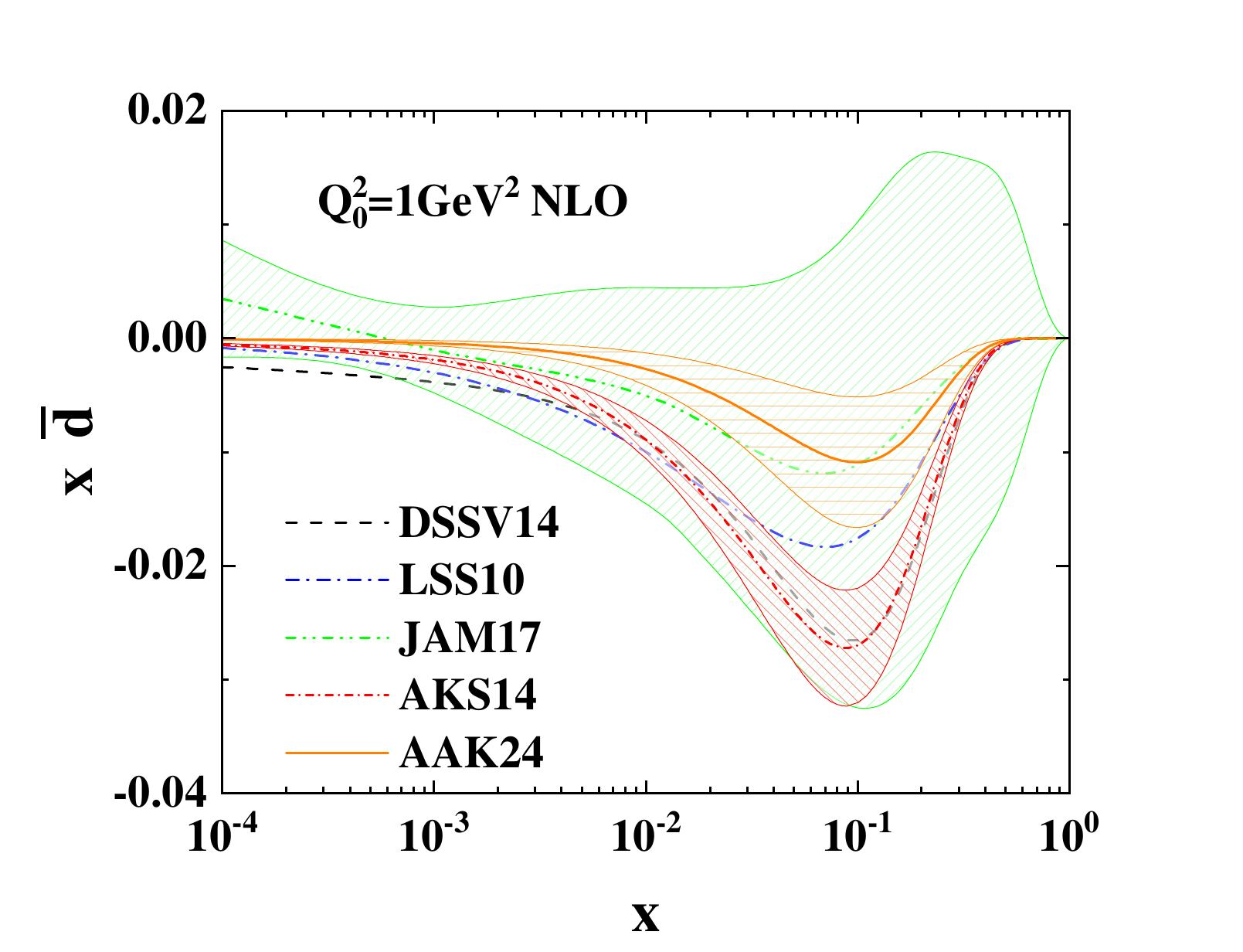}
\includegraphics[clip,width=0.32\textwidth]{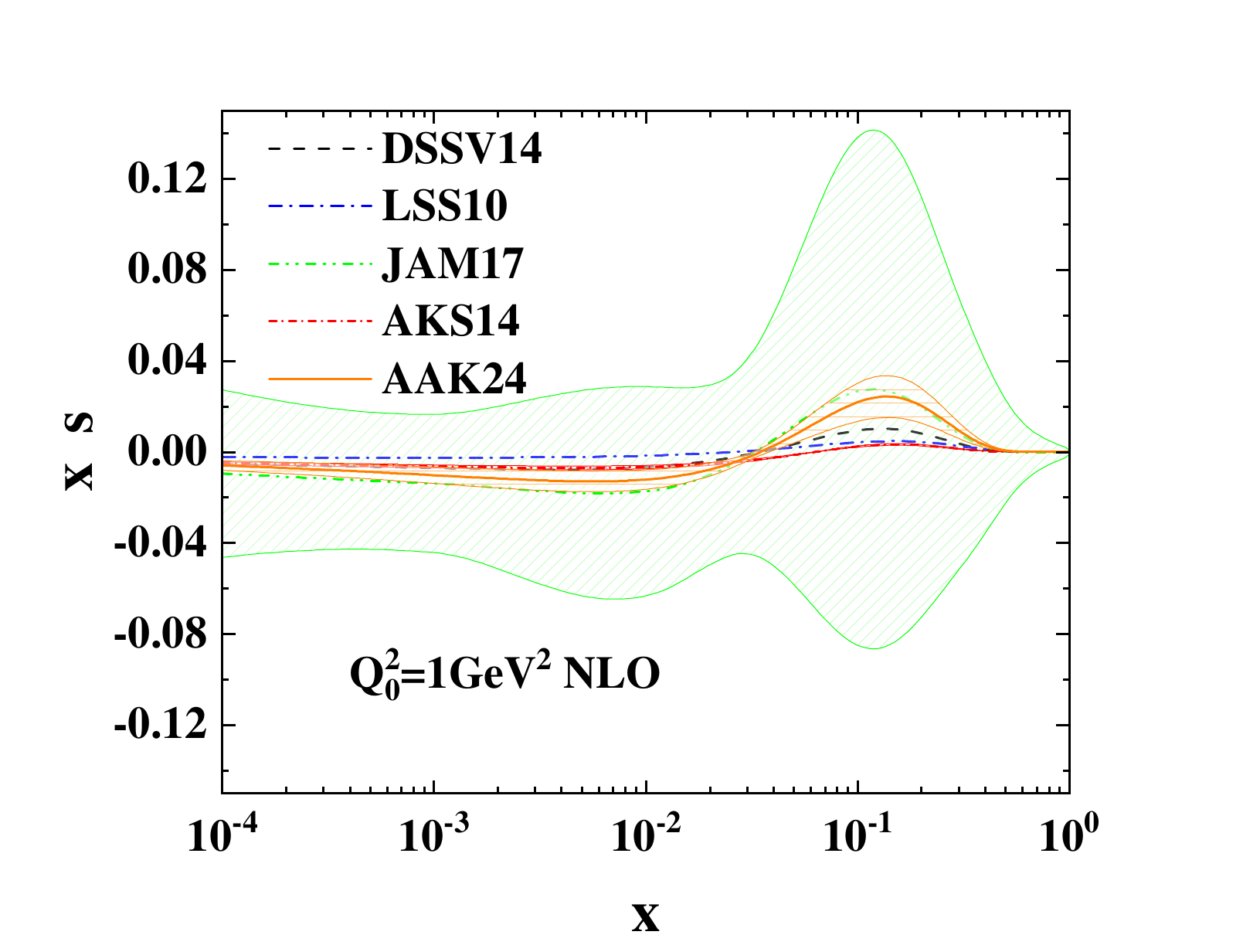}
\includegraphics[clip,width=0.32\textwidth]{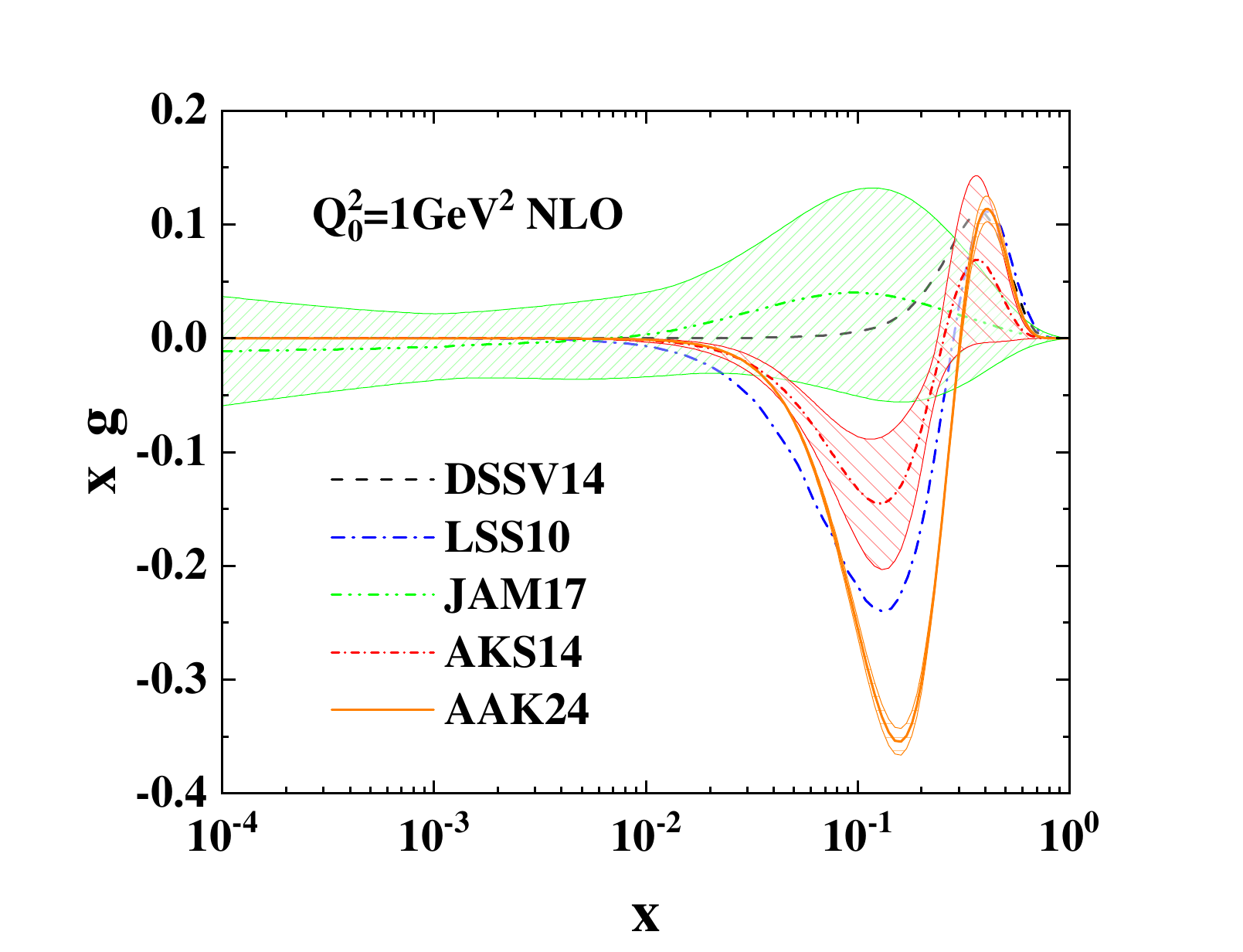}
\caption{The {\tt AAK24} quark and gluon helicity distributions at the input scale $Q^2_0=1$ GeV$^2$, with uncertainty 
bands now clearly corresponding to the different PDF sets. 
The comparisons include {\tt DSSV14}~\cite{deFlorian:2014yva} (black dashed), {\tt LSS10}~\cite{Leader:2010rb} (blue dashed dotted), 
{\tt JAM17}~\cite{Sato:2016tuz} (green dashed dot dotted), and {\tt AKS14}~\cite{Arbabifar:2013tma} (red short dashed dotted). 
The uncertainty bands for each set are shown with the same color as their corresponding line in the legend.}  
\label{fig4-AAK24parton_1}
\end{center}
\end{figure}

\begin{figure}
\begin{center}
\includegraphics[clip,width=0.32\textwidth]{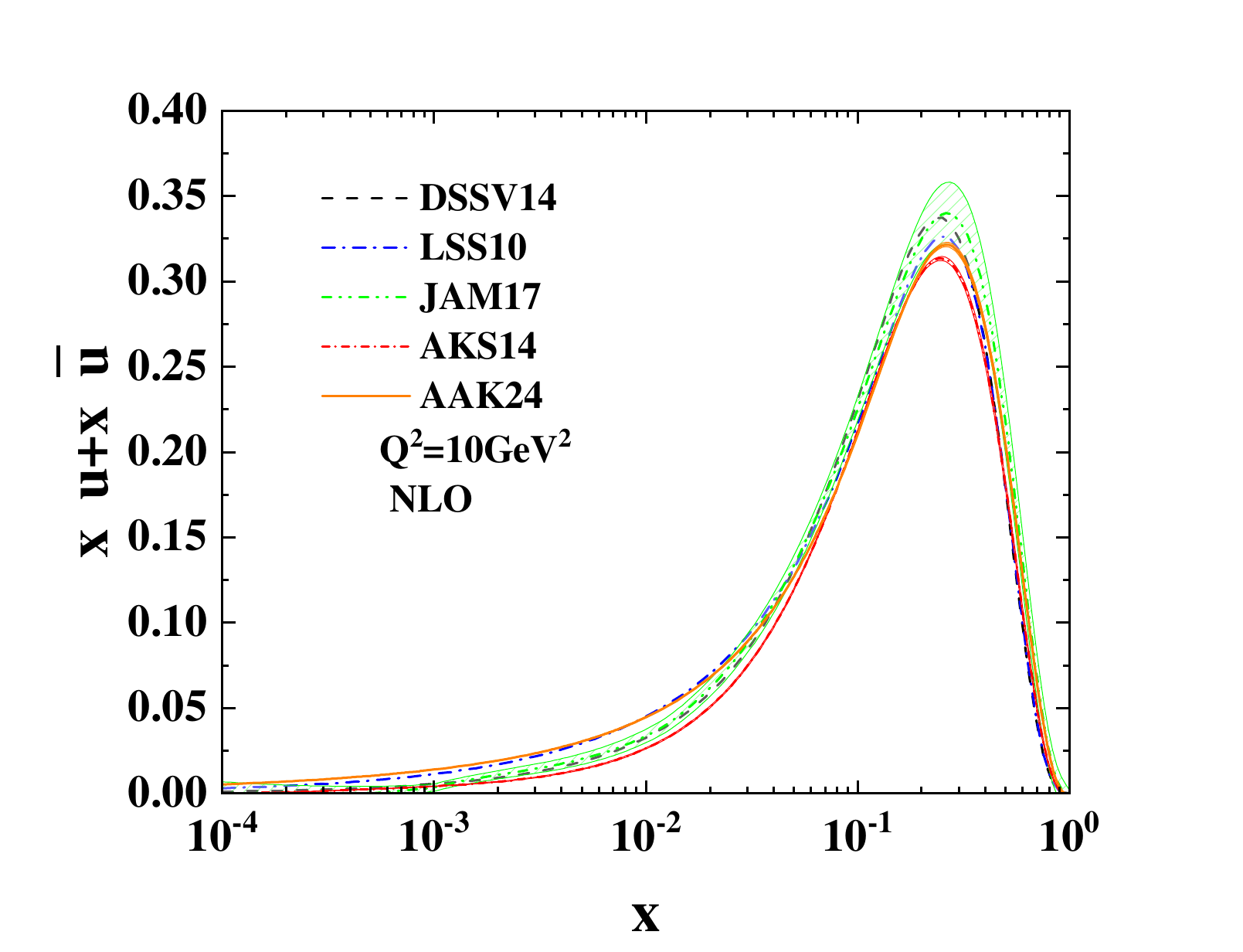}
\includegraphics[clip,width=0.32\textwidth]{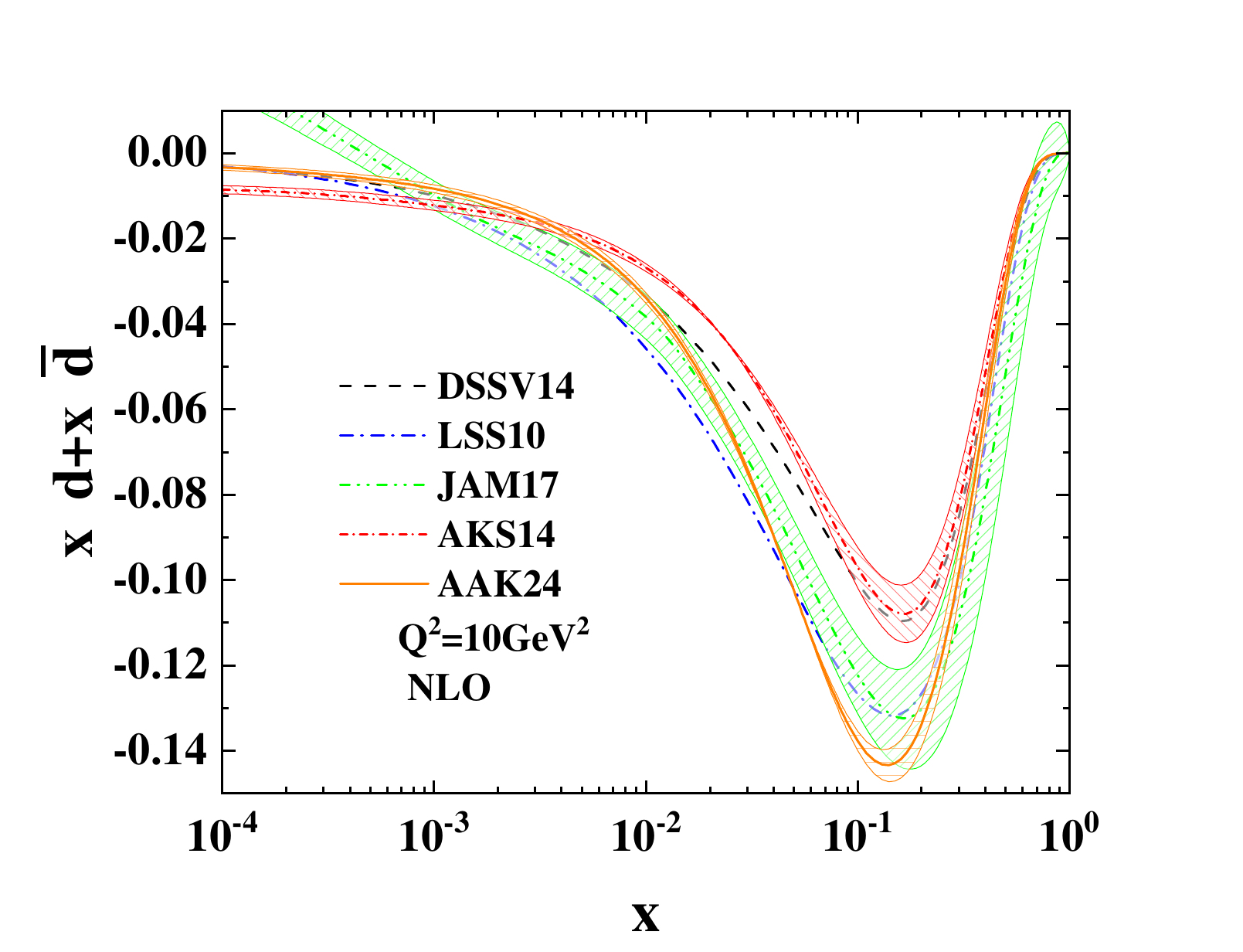}
\includegraphics[clip,width=0.32\textwidth]{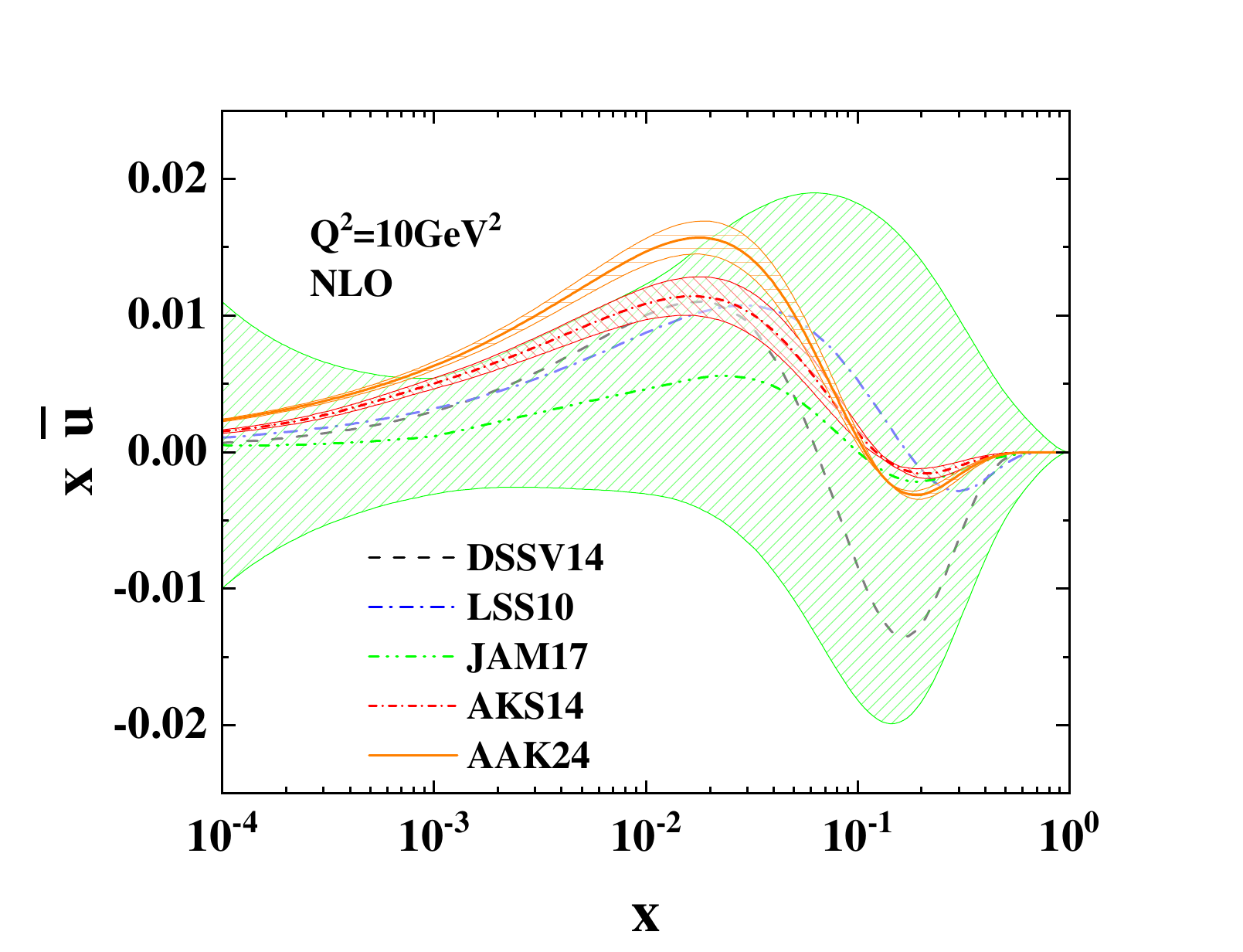}
\includegraphics[clip,width=0.32\textwidth]{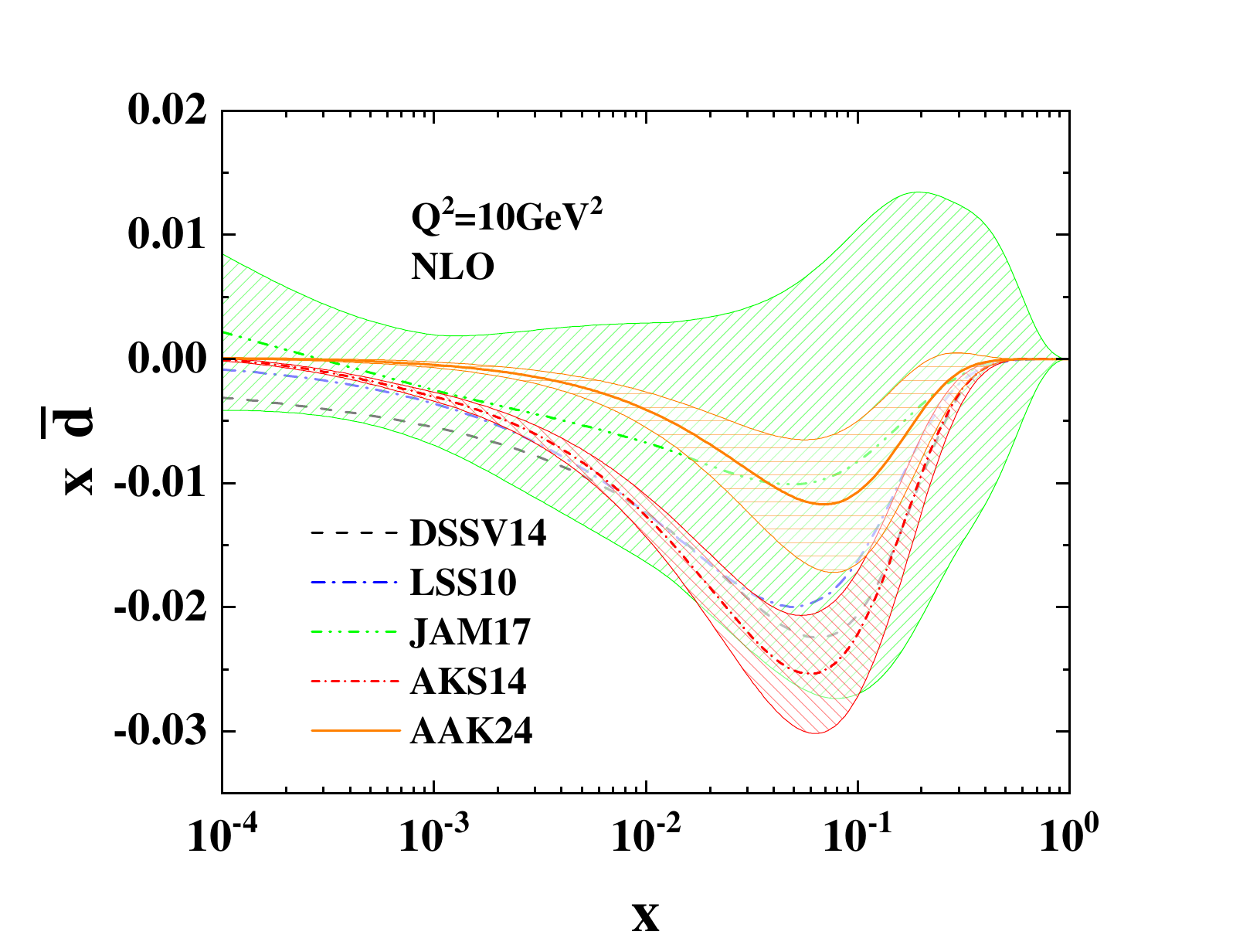}
\includegraphics[clip,width=0.32\textwidth]{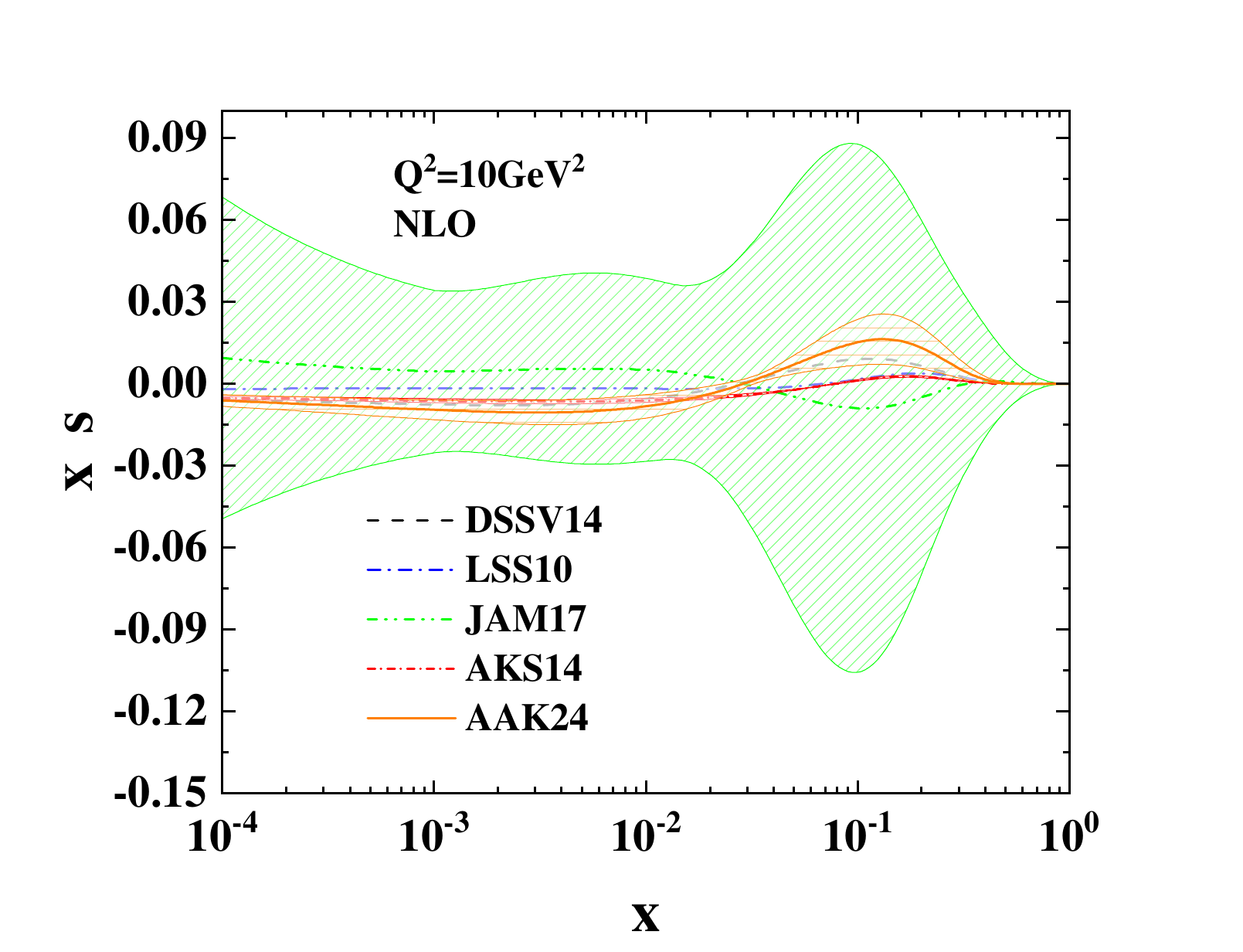}
\includegraphics[clip,width=0.32\textwidth]{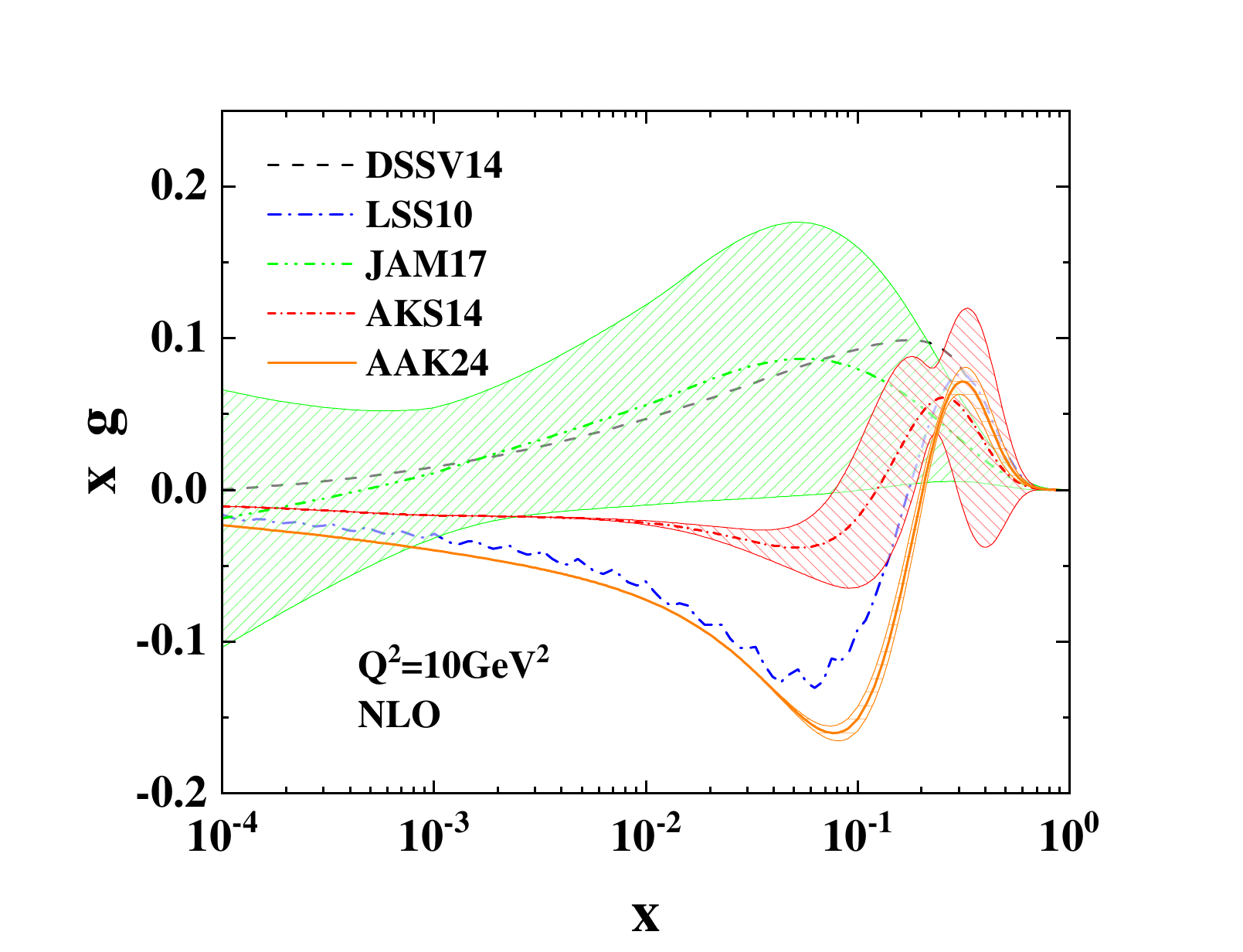}
\caption{Same as Fig.~\ref{fig4-AAK24parton_1} but this time at the higher value of 10 GeV$^2$.}
\label{fig4-AAK24parton_2}
\end{center}
\end{figure}

}

{ 
	
	%
	%
	\subsection{The AAK24 fit quality and the data/theory comparison} \label{data-theory-comparison}

	We will now shift our focus to discussing the fit quality of AAK24 and 
	conducting a detailed comparison of NLO theory predictions with the data analyzed in this study.
	
	The quality of the {\tt AAK24} global QCD analysis is 
	summarized in Tables.~\ref{table-data-A1}, 
	\ref{table-data-A2} and \ref{table-DIDIS-data}.
	For both the $A_1$ and $\frac{g_1}{F_1}$ observables, we have presented both the 
	total and individual $\chi^2$ values in Table~\ref{table-data-A1}. 
	It is noteworthy that in the total, we have considered 951 data points, resulting in a 
	total $\chi^2$ of 1170.172,  which indicates a relatively 
	good fit within the framework of QCD.
	When examining individual $\chi^2$ values, it becomes evident that 
	in most cases, there is a strong 
	agreement between the data and theory. 
	This is particularly evident in the case of the CLAS data, 
	which contributes significantly to 
	the total dataset with 136 data points for $A_1^p$ and 452 for $A_1^d$.
	For the newly added COMPASS and JLAB data, our observations indicate that 
	the fit quality is generally 
	good for COMPASS16 and COMPASS17. However, for the JLAB17 dataset, 
	a less favorable $\chi^2$ value is obtained, 
	suggesting a relatively worse agreement between the data and 
	theoretical predictions in this case.  
	
	The corresponding $\chi^2$  for the $A_2$ asymmetry observables, 
	with a total of $N_{dat} = 140$ data points, 
	are available in Table.~\ref{table-data-A2}. 
	It is worth noting that not only do we achieve a very favorable total $\chi^2$ value, 
	but we also observe good individual $\chi^2$ values for datasets originating from 
	various experiments analyzed in this work. 
	This suggests a strong agreement between the data and theoretical 
	predictions across different experimental sources.
	
	Finally, Table.~\ref{table-DIDIS-data} provides the $\chi^2$ values 
	for the $A_{1N}^h$ SIDIS observables that 
	have been examined in this study as well, encompassing a total 
	of $N_{dat} = 160$ data points. 
	As is evident from this table, the individual $\chi^2$ values for 
	each experiment exhibit a consistent and stable pattern, 
	and the overall total values indicate a good agreement between the 
	data and the theoretical predictions. 
	
	As discussed earlier, an additional analysis was performed where the $A_2$ asymmetry 
	observables (data in Table.~\ref{table-data-A2}) were excluded from the QCD fit. 
	The corresponding $\chi^2$ values per data point are presented in the last column of 
	Tables.~\ref{table-data-A1} and \ref{table-DIDIS-data}. The numbers in these tables 
	suggest that the exclusion of $A_2$ asymmetry data could potentially 
	increase the total values of $\chi^2$.
	Table.~\ref{table-data-A1} shows that the inclusion of $A_2$ data increases 
	the total value of $\chi^2$ from 1170.172 to 1304.509. Similarly, Table.~\ref{table-DIDIS-data} 
	indicates an increase in the total value of $\chi^2$ from 227.934 to 244.520.

	In summary, the global $\chi^2/N_{dat}$ value for the entire 
	dataset of 1266 data points in the {\tt AAK24} 
	global QCD analysis is calculated to be 1.0528. 
	This value indicates a stable and acceptable fit quality, 
	demonstrating that the theoretical model 
	aligns well with the experimental data.
	However, when the $A_2$ asymmetry observables are excluded, 
	the $\chi^2/N_{dat}$ value increases to 1.0647. 
	This result underscores the significance of including the $A_2$ 
	asymmetry data in QCD analyses of polarized PDFs. 
	It suggests that the inclusion of $A_2$ data leads to a 
	better-constrained set of polarized PDFs with 
	improved agreement between the data and theory, emphasizing the 
	importance of these measurements in 
	refining our understanding of polarized PDFs.

	We can now proceed to compare the NLO theory predictions obtained for the 
	asymmetry observables used in this study with the experimental datasets, 
	particularly the newly added polarized data from {\tt COMPASS16}~\cite{COMPASS:2015mhb} 
	and {\tt COMPASS17}~\cite{COMPASS:2016jwv} experiments, 
	focusing on the $A_1^p$ and $A_1^d$ asymmetries.
	Such comparisons are displayed in Figs.~\ref{fig6-A1pcompass16} and ~\ref{fig7-A1dcompass17} 
	for the $A_1^p$ and $A_1^d$ asymmetry data from the {\tt COMPASS16} 
	and {\tt COMPASS17} experiments, respectively. 
	The comparisons are made as functions of $x$ for two different 
	selections of $Q^2$, specifically 15 and 13 GeV$^2$.
	As these plots illustrate, the predictions from the {\tt AAK24} 
	polarized PDFs are in excellent 
	agreement with both the {\tt COMPASS16} and {\tt COMPASS17} datasets across 
	the entire range of $x$. 
	These results align with the individual $\chi^2$ values 
	presented in Table.~\ref{table-data-A1}, 
	further confirming the quality of the fit and the compatibility 
	between the theoretical model 
	and the experimental data.

In Figs.~\ref{fig3-2} and \ref{fig3-3}, we present the polarized proton 
structure functions \(xg^p_2\) calculated using the Wandzura-Wilczek relation in Eq.~\ref{g2-ddis}. 
The results from {\tt  NMA23} (dashed dashed dotted)~\cite{Nematollahi:2023dvj} and 
{\tt  KTA17} (dashed dotted)~\cite{Khanpour:2017cha} are also shown. The experimental 
data from the E155 collaboration~\cite{E155:2002iec} are included for comparison. 
As can be seen, the \(xg^p_2\) structure functions calculated using the 
Wandzura-Wilczek approximation provide good agreement with both the data and 
other results available in the literature.

	To provide additional support for our discussion regarding 
	the comparison between data and theory, 
	we present detailed comparisons of our 
	theoretical predictions with a 
	different selection of $A_1^p$ asymmetry data as outlined 
	in Table.~\ref{table-data-A1}.
	As one can see from Fig.~\ref{A1}, the agreement between the theoretical model and 
	various selections of $A_1^p$ asymmetry data is good.
	Such agreement between the theoretical predictions and a 
	variety of experimental datasets 
	for $A_1^p$ asymmetry provides evidence of the validity and 
	reliability of the {\tt AAK24} analysis in describing the 
	polarized PDFs in nucleons.

\begin{figure}
	\begin{center}
		\includegraphics[clip,width=0.45\textwidth]{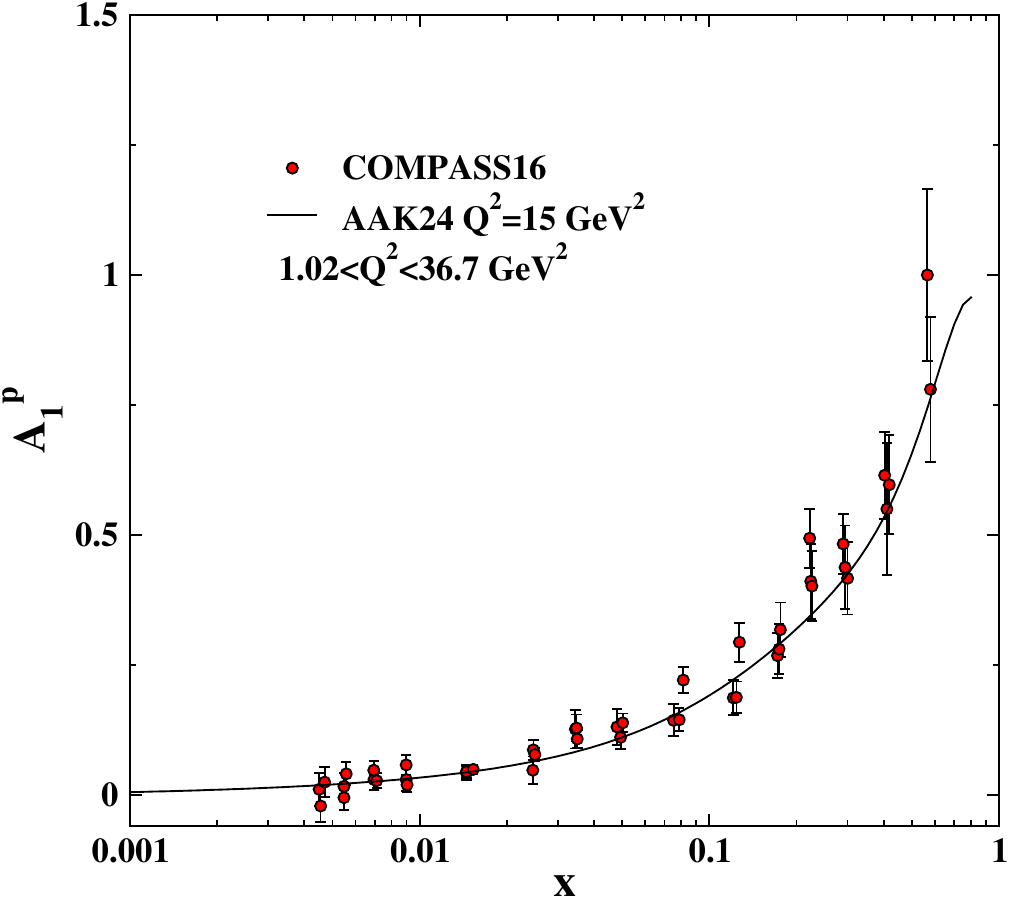}
		\caption{Comparison of the {\tt AAK24} theory prediction with the $A_1^p$ 
			asymmetry data from {\tt COMPASS16} experiment~\cite{COMPASS:2015mhb}.}
		\label{fig6-A1pcompass16}
	\end{center}
\end{figure}

\begin{figure}
	\begin{center}
		\includegraphics[clip,width=0.45\textwidth]{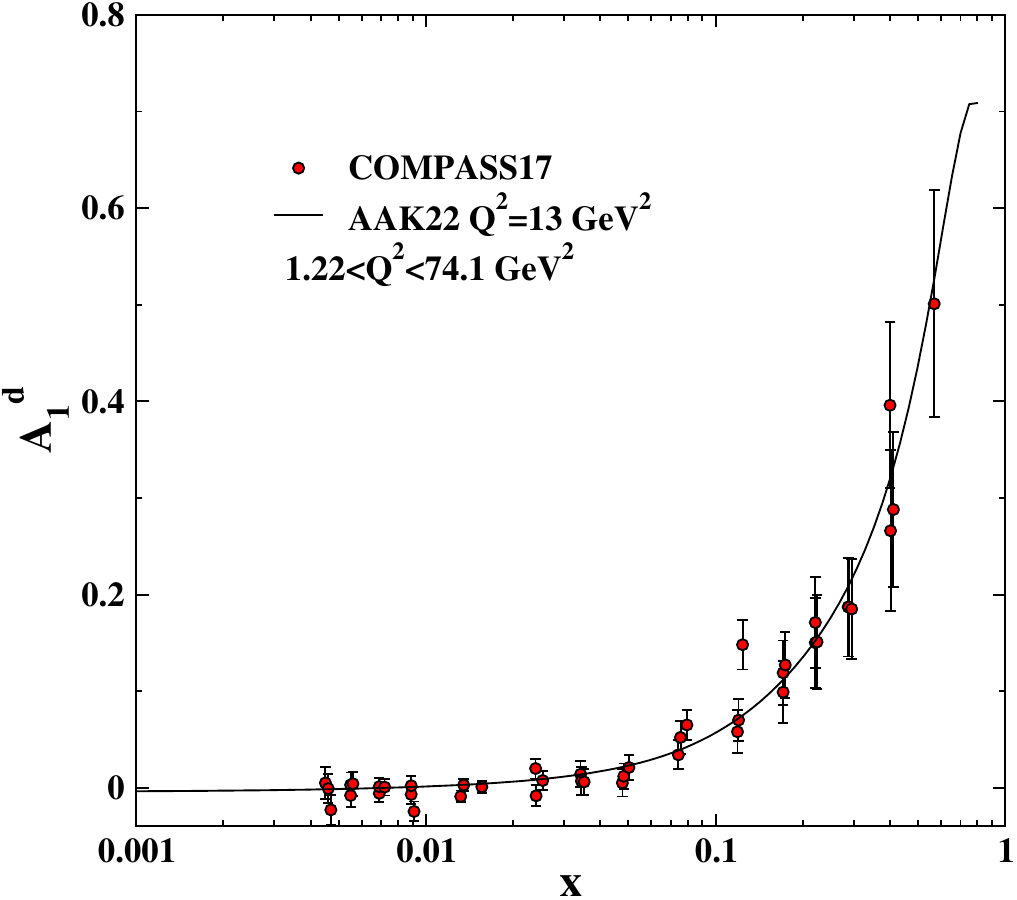}
		\caption{Same as Fig.~\ref{fig6-A1pcompass16} but this time for the $A_1^d$ asymmetry 
			data from {\tt COMPASS17} experiment~\cite{COMPASS:2016jwv}.}
		\label{fig7-A1dcompass17}
	\end{center}
\end{figure}

\begin{figure}
	\begin{center}
		\includegraphics[clip,width=0.55\textwidth]{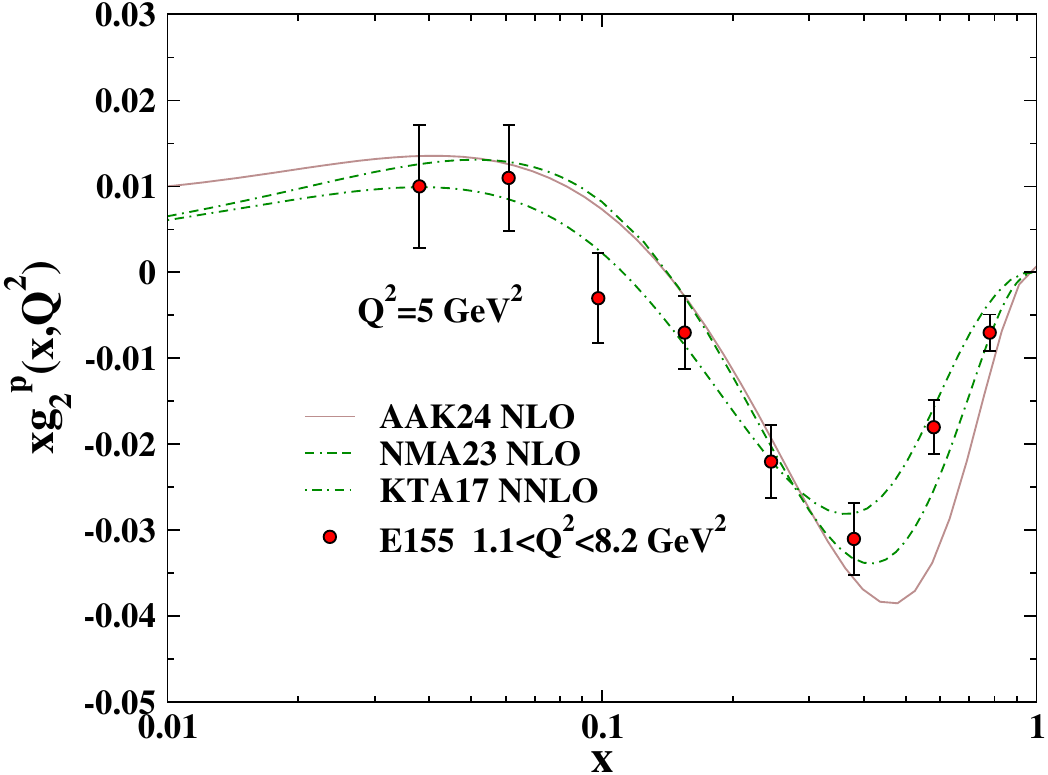}
		\caption{The polarized proton structure functions $xg^p_2$ as a function of $x$ at $Q^2=5$ GeV$^2$. 
			The results from {\tt NMA23} (dashed dashed dotted)~\cite{Nematollahi:2023dvj} 
			and {\tt  KTA17} (dashed dotted)~\cite{Khanpour:2017cha} are also shown. 
			The experimental data from the E155 collaboration~\cite{E155:2002iec} are presented for comparison.}
		\label{fig3-2}
	\end{center}
\end{figure}

\begin{figure}
	\begin{center}
		\includegraphics[clip,width=0.55\textwidth]{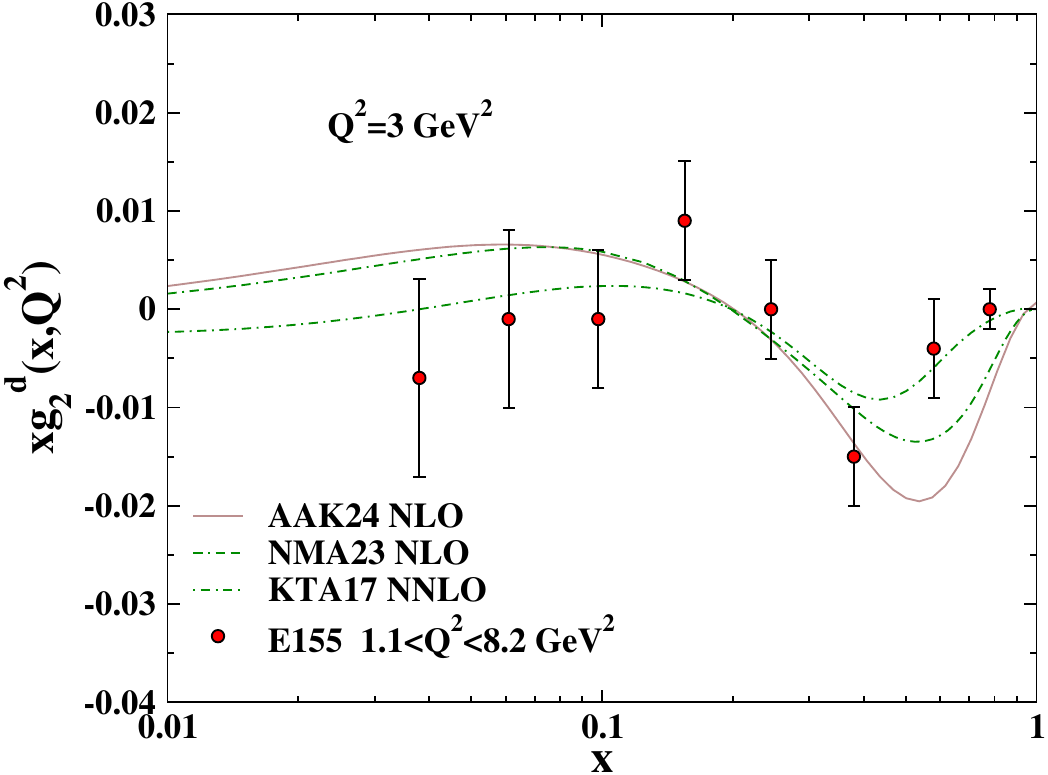}
		\caption{Same as Fig.~\ref{fig3-2}, but this time the comparisons are shown for $Q^2$=3 GeV$^2$. } 
			\label{fig3-3}
	\end{center}
\end{figure}

\begin{figure}
	\begin{center}
		\includegraphics[clip,width=0.8\textwidthclip,width=0.8\textwidth]{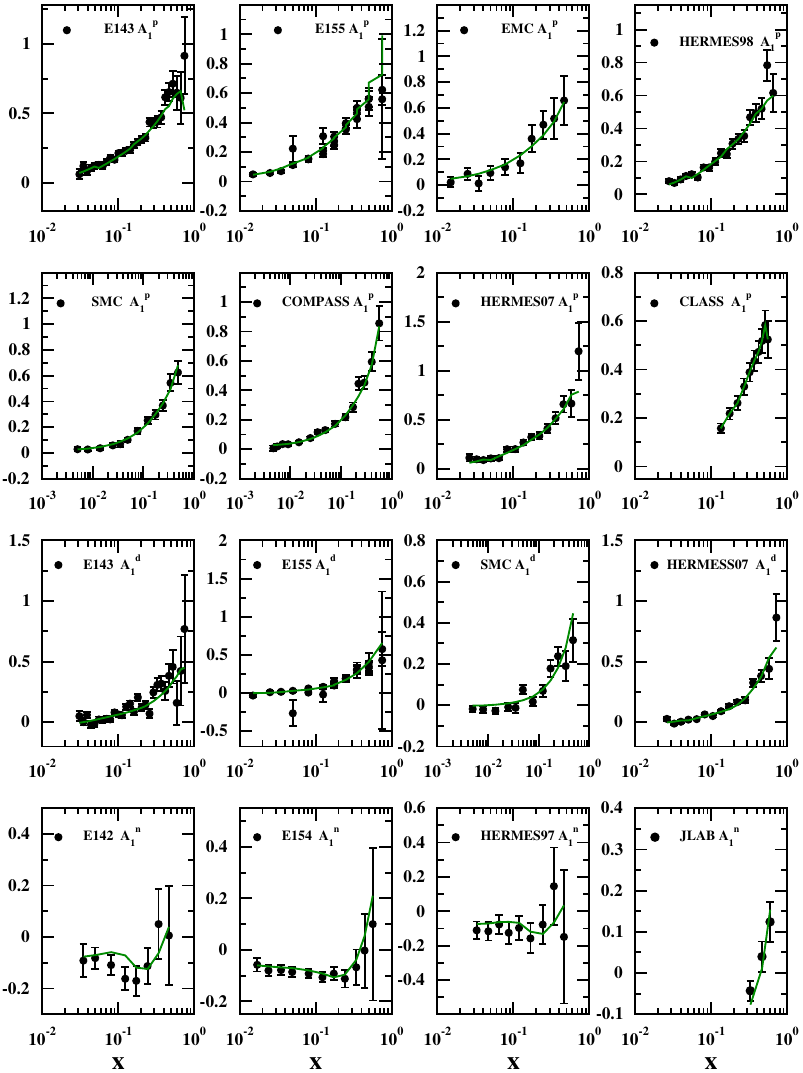}
		\caption{Comparison of the {\tt AAK24} theory prediction with 
			some selection of $A_1^p$, $A_1^n$, and $A_1^d$ 
			asymmetry data presented in Table.~\ref{table-data-A1}.}
		\label{A1}
	\end{center}
\end{figure}

	We are now in a position to compare the theoretical predictions from {\tt AAK24} with the 
	corresponding experimental results for the SIDIS asymmetries. 
	This comparison is illustrated in Fig.~\ref{A1SIDIS} for selected 
	data from SMC and COMPASS collaborations.  
	Our calculations include the theory predictions for proton, neutron, 
	and deuteron, taking into account 
	the light charged hadron, kaon, and pion FFs.
	As shown in the figures, the agreement between the data and theory is excellent across 
	the entire range of $x$ and for all hadron species. 
	This alignment corroborates the favorable $\chi^2$ values 
	presented in Table.~\ref{table-DIDIS-data}. 
	It further confirms the consistency between the theoretical framework and the 
	experimental data, affirming the model ability to accurately describe the 
	observed phenomena in the realm of polarized SIDIS asymmetries.

	\begin{figure}
		\begin{center}
		\includegraphics[clip,width=0.8\textwidthclip,width=0.8\textwidth]{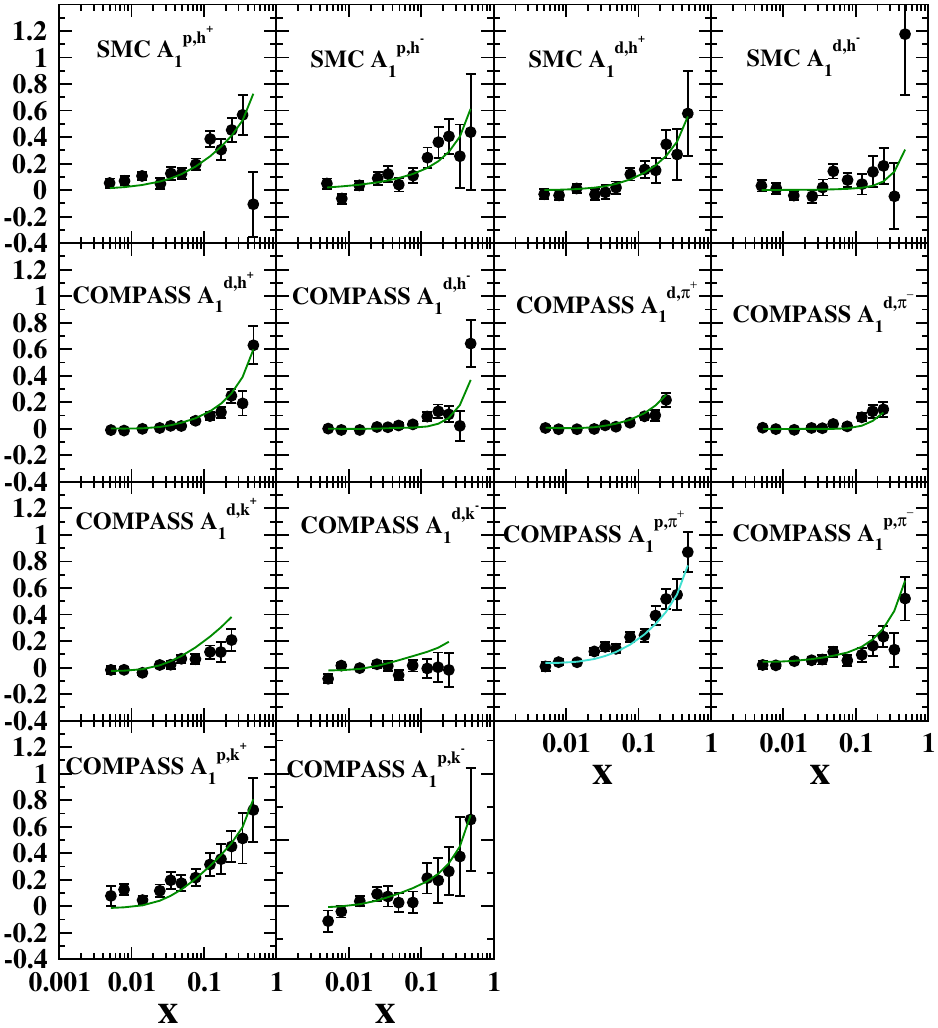}
		\caption{Comparison of {\tt AAK24} QCD results for the SIDIS asymmetries with 
			the corresponding SIDIS observables. 
			The calculations include the theory predictions for proton, neutron, and deuteron, 
			taking into account the light charged hadron, kaon, and pion FFs.}
		\label{A1SIDIS}
\end{center}
\end{figure}
	
	To provide more detailed discussions of the results, we provide a comparison in 
	Fig.~\ref{A2DIS} between the {\tt AAK24} theoretical predictions for 
	the $A_2$ asymmetry for proton, neutron, and deuteron targets and the 
	corresponding data analyzed in this study. 
	As mentioned earlier, this marks the first inclusion of $A_2$ asymmetry data in 
	a global QCD analysis of polarized PDFs.
	Upon inspection, the data/theory agreement is generally good for most experiments. 
	However, some deviations between the data and theory are evident. For example, 
	discrepancies can be observed 
	for the $A_2^d$ asymmetry from the SMC experiment and 
	the $A_2^p$ asymmetry from HERMES. 
	
	\begin{figure*}[h]
			\begin{center}
		\includegraphics[clip,width=0.8\textwidthclip,width=0.8\textwidth]{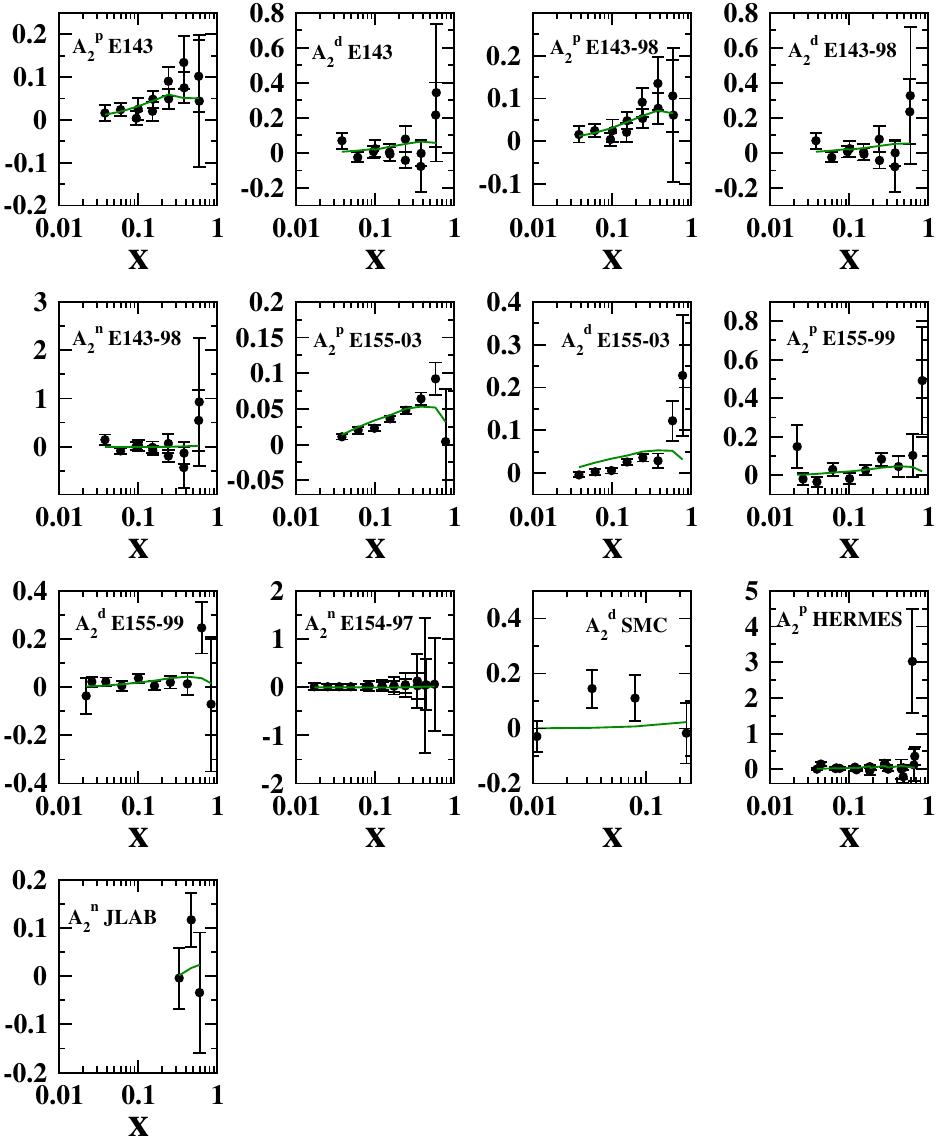}
		\caption{Comparison of {\tt AAK24} QCD results for the $A_2$ asymmetries 
			with a selected set of corresponding data. 
			The calculations include the theory predictions for proton, neutron, and deuteron}
		\label{A2DIS}
			\end{center}
	\end{figure*}

In summary, the {\tt AAK24} results presented in this section, which integrated recent 
experimental data from {\tt JLAB17}, {\tt COMPASS16}, and {\tt COMPASS17}, including the 
novel inclusion of $A_2$ asymmetry measurements, exhibited favorable overall agreement 
with experimental data. 
These findings also align well with other polarized PDF determinations, 
notably {\tt JAM17}, {\tt DSSV14}, {\tt LSS10}, and {\tt AKS14}, all of which 
also accounted for SU(2) and SU(3) symmetry breaking.

}

{

	%
	%
	\section{Conclusions}\label{Conclusions}

	In this article, we have presented {\tt AAK24} polarized PDFs, 
	a next-to-leading order (NLO) QCD analysis 
	of polarized data from both polarize DIS 
	and SIDIS experiments on the nucleon. 
	Throughout the analysis, we considered the scenario of SU(2) and SU(3) 
	symmetry breaking, specifically $\delta \bar{u} \neq \delta \bar{d} \neq \delta \bar{s}$, 
	while treating $\delta \bar{s}$ and $\delta s$ as equal since the available experimental
	data are not enough to distinguish them. 
Additionally, we have included the nonperturbative target mass corrections (TMC) as 
		well as higher-twist terms (HT) which are particularly important for the polarized DIS. 
	Our study addressed the crucial role of semi-inclusive data in 
	determining the polarized sea quark distributions.
	The recent experimental data from {\tt JLAB17}, {\tt COMPASS16}, and 
	{\tt COMPASS17} experiments have 
	been incorporated into our data samples. 
	Additionally, this analysis marks simultaneous incorporation of SIDIS and  
	$A_{2}$ asymmetry measurements as well. 
	
	We have thoroughly examined the impact of these new data sets on 
	both the central values and the extracted uncertainties, 
	as well as their effect on the 
	overall fit quality. 
	Our findings are presented and discussed to provide a comprehensive 
	understanding of their implications.
	To quantify the uncertainties in the polarized PDFs, we employed the 
	standard Hessian method. 
	The main results of {\tt AAK24} QCD analysis exhibit an overall 
	good agreement with the analyzed experimental data, 
	aligning well with other polarized PDFs determinations which considered the SU(2) and SU(3) 
	symmetry breaking as well. 
	
	A Fortran package containing {\tt AAK24} polarized PDFs at NLO accuracy 
	can be obtained via e-mail from the authors. 
	
}

{ 
	\section*{Data Availability Statement} 
	This manuscript has associated data in a data in a repository. 
	[Authors comment: This data sets generated during the current study are available from the corresponding author on responsible request.] 
	
	%
	%
	\section*{Acknowledgments}

	The authors appreciate the financial support from the 
	Iran National Science Foundation (INSF) under grant number 4013570. 
	S. A. T. and H. K. are also grateful to the School of Particles and Accelerators, 
	Institute for Research in Fundamental Sciences (IPM). 
	F.~A. acknowledges the Farhangian University for the provided support to conduct this research. 
	H. Khanpour appreciates the financial support from NAWA
	under grant number BPN/ULM/2023/1/00160 and from the IDUB programme
	at the AGH University.
	We also thank Dimiter Stamenov for providing the necessary {\tt LSS10} grids, and 
	Rodolfo Sassot for supplying the required {\tt DSSV14} grids data.
	
}

\clearpage
%
%

\end{document}